\documentclass[fleqn,10pt]{wlscirep}
\usepackage[utf8]{inputenc}
\usepackage[percent]{overpic}
\usepackage[T1]{fontenc}
\usepackage{url}
\usepackage{hyperref}

\title{Biswas-Chatterjee-Sen (BChS) kinetic exchange opinion model on modular networks}

\author[1]{Hrishidev Unni}
\author[2]{Soumyajyoti Biswas}
\author[1,*]{Anirban Chakraborti}
\affil[1]{School of Computational \& Integrative Sciences, Jawaharlal Nehru University, New Delhi-110067, India}
\affil[2]{Department of Physics, SRM University - AP, Andhra Pradesh 522240, India}

\affil[*]{anirban@jnu.ac.in}

\begin{abstract}
We study opinion formation in a society where agents interact on a modular network generated using a stochastic block model (SBM). Opinion dynamics is modeled through the Biswas-Chatterjee-Sen (BChS) kinetic exchange model, in which agents undergo pairwise interactions that could be positive or negative. By tuning the relative strength of intra- and inter-group connectivity inherent to the SBM, as well as the disagreement probability, we identify distinct collective phases. In particular, we observe a robust regime with strong intragroup ordering but no global consensus, in addition to fully ordered and disordered states. In the particular case of two modules, we observe an anti-ferromagnetic type ordering with the increase of negative interaction between the groups. We show approximate analytical calculations and numerical results of it. These results demonstrate how modular interaction structure can qualitatively alter collective opinion dynamics and hinder consensus formation.
\end{abstract}
\begin{document}

\flushbottom
\maketitle
% * <john.hammersley@gmail.com> 2015-02-09T12:07:31.197Z:
%
%  Click the title above to edit the author information and abstract
%
\thispagestyle{empty}

% \noindent Please note: Abbreviations should be introduced at the first mention in the main text - no abbreviations lists. Suggested structure of main text (not enforced) is provided below.

\section{Introduction}

Whether a collection of interacting humans can be viewed in a similar manner as a collection of colliding atoms is a question that has intrigued social scientists and physicists for a long time~\cite{Castellano_RMP_2009,Sen_Chakrabarti_Sociophysics_2013,Galam_Sociophysics_2012}. While the obvious arguments against this approach remain the oversimplification of nuances in human emotions to mere numbers, the proposition in favour is the possibility of an exploration-similar in spirit to critical phenomena-providing a universal description of collective behaviour arising out of such interactions~\cite{Castellano_RMP_2009}. To many, this line of exploration is the only pathway for such descriptions to evolve towards a predictive, falsifiable, and reproducible science~\cite{Sen_Chakrabarti_Sociophysics_2013,Galam_Sociophysics_2012}.

Even if the premise of collective behaviour (hence eliminating complications from individual human characteristics) is accepted, there are obvious hurdles on the way towards a mathematical description of human interactions and the resulting opinion formation. These questions include, for example, the quantification of opinions, setting up rules of interactions, and identification of groups or topologies within which such interacting individuals live~\cite{Castellano_RMP_2009}. For quantification of opinions, it is possible to restrict to situations where it is binary (or at least limited within a finite set of values), say for a movie rating, voting in countries with two major parties, or voting in a referendum~\cite{Castellano_RMP_2009}. Setting up rules of interactions is a more complicated question. However, as indicated earlier, we reduce the interacting population to a series of binary interactions, as in the case of the kinetic theory of an ideal gas~\cite{Lallouache_PRE_2010,BChS_frontiers_2023}. In such cases, unlike the case of an ideal gas or that of wealth exchange models, there is no requirement of any conservation constraint~\cite{Chatterjee_Chakrabarti_EPJB_2007,Chakraborti_EconophysicsReviewI_2011,Chakraborti_EconophysicsReviewII_2011,Chatterjee_EconophysicsWealth_2005,Sinha_EconophysicsIntro_2010,Chakrabarti_CUP_IncomeWealth_2013,Abergel_NewPerspectivesEconophysics_2019}. This would mean that even starting from a state of fragmented opinion values, the society as a whole can evolve towards a consensus through these binary exchange interactions, following certain parameter values embedded in the interaction rules~\cite{Lallouache_PRE_2010,Biswas_PRE_2011_MF,biswas2012disorder,BChS_frontiers_2023}. A recent review on kinetic-exchange-type models for opinion dynamics can be found in Ref.~\cite{BChS_frontiers_2023}; other reviews of such dynamics and models include Refs.~\cite{Castellano_RMP_2009,Sen_Chakrabarti_Sociophysics_2013}.

However, in most cases, the kinetic exchange opinion models are studied either in fully connected graphs (mean-field models) or in some regular lattice~\cite{Lallouache_PRE_2010,Biswas_PRE_2011_MF,biswas2012disorder,BChS_frontiers_2023}. In some cases, other types of networks were also studied~\cite{LIMA2021125834,Pranesh_Gupta_2025}. From a complex systems perspective, collective behaviour is shaped not only by microscopic interaction rules but also by the mesoscopic structure of the interaction network itself~\cite{Chakrabarti_DataScienceComplexSystems_2023,Kozma_Barrat_PRE_2008}. Changes in connectivity patterns can qualitatively alter macroscopic phases and transitions, even when the underlying dynamical rules remain unchanged~\cite{Kozma_Barrat_PRE_2008}.

Here, we focus on the interaction network, especially when it acquires a modular structure. This means that there are groups of individuals who interact more closely among themselves than between members of such other groups~\cite{Newman_Modularity_PNAS_2006,newman2010networks}. This is indeed a rather realistic feature that conforms with the ideas of echo chambers, where in an increasingly polarised world, there is often very little appetite to exchange ideas with members of different groups~\cite{Artime_PRL_2020,Echo_Vaccine_BMC_2022,BorgeHolthoefer_SciRep_2017,Contrarian_Echo_2019}. This can potentially give rise to a lack of consensus, even in the parameter regions where such consensus is likely to form in a uniformly connected (non-modular) environment~\cite{Kozma_Barrat_PRE_2008,BorgeHolthoefer_SciRep_2017,Mucha_Science_2010}.

In this work, we investigate the fragility of consensus formation in modular networks using the kinetic exchange (BChS) model of opinion dynamics~\cite{biswas2012disorder,BChS_frontiers_2023}. A closely related study examined the BChS model on a modular interaction structure consisting of two groups, highlighting how inter-group coupling influences collective ordering~\cite{Suchecki2group}. Here, we extend this framework to stochastic block networks with multiple groups and tunable connectivity~\cite{holland1983stochastic,KarrerNewman_SBM_2011}. We look at the phases and transitions seen in the model with varying degree of modularity and disagreement probability~\cite{biswas2012disorder,BChS_frontiers_2023}.

Complementary to Monte Carlo simulations of the BChS microscopic update rule on SBMs, we also use a group-level mean-field (ODE) formulation to interpret the phase structure in terms of mixing eigenmodes and to expose initial-condition dependence in modular settings. Related three-state kinetic-exchange dynamics on modular (community-structured) networks, including inter-group negative interactions and community-preserving mean-field closures, have been analysed in Ref.~\cite{Oestereich_2019}. More broadly, recent mechanistic agent-based work has highlighted how platform-mediated information-spreading mechanisms can co-evolve with opinion polarisation and echo-chamber formation, motivating a network-structure-aware treatment of consensus fragility in online-like interaction environments~\cite{oliveira2025oppol}.

%----------------------------------------------------------------------------------------
%	MODEL AND SIMULATIONS
%----------------------------------------------------------------------------------------
\section{Model and Simulations}
\label{sec:ch5_model}

\subsection{Stochastic Block Networks}
\label{subsec:sbm_networks}
Interaction networks are generated using a stochastic block model (SBM) with equal-sized groups. The model is specified by four parameters: the total number of agents $n$, the number of groups $c$, the within-group link probability $p_{\mathrm{in}}$, and the between-group link probability $p_{\mathrm{out}}$. The $n$ agents are partitioned into $c$ disjoint modules of size $n/c$. For any unordered pair of agents $(i,j)$,
\begin{itemize}
    \item if $i$ and $j$ belong to the same module, an undirected edge is present with probability $p_{\mathrm{in}}$;
    \item if they belong to different modules, an undirected edge is present with probability $p_{\mathrm{out}}$.
\end{itemize}
All edges are drawn independently. Throughout most of the study we fix a relatively large $p_{\mathrm{in}}$ and vary $p_{\mathrm{out}}$ to control the strength of the modular structure: small $p_{\mathrm{out}}$ corresponds to strongly segregated communities, whereas larger $p_{\mathrm{out}}$ interpolates toward a well-mixed network. In a separate set of scans we also vary $p_{\mathrm{in}}$ and $p_{\mathrm{out}}$ jointly to explore how the phase diagram depends on the absolute densities of intra- and inter-group connections.

\subsection{BChS Opinion Dynamics}
\label{subsec:bchs_dynamics}
On a given network, opinions evolve according to the three-state Biswas--Chatterjee--Sen (BChS) kinetic-exchange model. Each agent $i$ carries an opinion $o_i(t) \in \{-1,0,+1\}$ at discrete time $t$. At each update step:
\begin{enumerate}
    \item An edge $(i,j)$ is selected uniformly at random from the edge list of the SBM.
    \item With probability $1/2$ agent $i$ updates based on $j$, and with probability $1/2$ agent $j$ updates based on $i$. Denoting the influenced agent by $a$ and the influencer by $b$, the opinion update is
    \begin{equation}
        \label{eq:bchs_update}
        o_a(t+1) = \mathrm{clip}\bigl(o_a(t) + \mu\, o_b(t)\bigr),
    \end{equation}
    where $\mu = +1$ (attractive interaction) with probability $1-p$ and $\mu = -1$ (repulsive interaction) with probability $p$, and $\mathrm{clip}(x)$ projects back to the allowed set $\{-1,0,+1\}$.
    \item All other opinions remain unchanged during that step.
\end{enumerate}
Here $p$ is the disagreement probability that controls the relative weight of repulsive versus attractive exchanges. Time is measured in units of single-edge updates.

\subsubsection*{Order parameters}
\label{subsec:order_parameters}
We monitor two order parameters. The global order parameter is the absolute mean opinion
\begin{equation}
    \label{eq:O_def}
    O(t) = \biggl|\frac{1}{n}\sum_{i=1}^n o_i(t)\biggr|,
\end{equation}
which measures net global alignment. To quantify modular ordering, we compute, for each module $g=1,\dots,c$, its local mean opinion
\begin{equation}
    \label{eq:mg_def}
    m_g(t) = \frac{1}{n/c} \sum_{i\in g} o_i(t),
\end{equation}
and define the intra-group order parameter as the average absolute module opinion
\begin{equation}
    \label{eq:Ointra_def}
    O_{\mathrm{intra}}(t) = \frac{1}{c} \sum_{g=1}^c |m_g(t)| .
\end{equation}
Large $O$ and large $O_{\mathrm{intra}}$ correspond to global consensus; large $O_{\mathrm{intra}}$ but small $O$ indicate a modularly polarised state with internally ordered but mutually opposed groups and small $O$ and small $O_{\mathrm{intra}}$ correspond to a fully disordered state.

\subsection{Monte Carlo Simulations}
\label{subsec:simulation_protocol}
For each choice of $(n,c,p_{\mathrm{in}},p_{\mathrm{out}},p)$ we proceed as follows. Agents' initial opinions are drawn independently and uniformly from $\{-1,0,+1\}$. The BChS dynamics is then iterated for $T_{\mathrm{s}} = 2\times 10^{6}$ single-edge updates to allow the system to reach a stationary regime. In the subsequent measurement phase of length $T_{\mathrm{m}} = 2\times 10^{6}$ updates we record $O(t)$ and $O_{\mathrm{intra}}(t)$ and compute their time averages,
\begin{equation}
    \label{eq:time_avgs}
    \langle O \rangle_t = \frac{1}{T_{\mathrm{m}}}\sum_{t=1}^{T_{\mathrm{m}}} O(t),
    \qquad
    \langle O_{\mathrm{intra}} \rangle_t = \frac{1}{T_{\mathrm{m}}}\sum_{t=1}^{T_{\mathrm{m}}} O_{\mathrm{intra}}(t),
\end{equation}
as well as higher moments used for Binder-type cumulants.

To obtain ensemble averages, we repeat this procedure for many independent realisations. For each parameter combination we generate $nI$ independent SBM network realisations (with $nI$ chosen depending on the desired resolution of the phase diagram; the value of $nI$ is stated below each corresponding result). On each network, we run independent BChS trajectories. Reported values of $\langle O \rangle$ and $\langle O_{\mathrm{intra}} \rangle$ in the phase diagrams are averages over both time and this ensemble of network realisations.

\subsection{Mean-field Approach}
\subsubsection*{SBM Mixing Matrix}
\label{subsec:sbm_mapping}
The Monte Carlo phase diagrams in Section~\ref{sec:ch5_results} are obtained from stochastic
simulations of the microscopic update rule on SBM networks. To interpret these regimes and to
connect to deterministic ODE simulations, we use a group-level mean-field description based on
the SBM neighbour mixing matrix $\Pi=[\pi_{gh}]$, where $\pi_{gh}$ is the probability that a random
neighbour of a node in group $g$ lies in group $h$.

\noindent\textit{Remark.} The Monte Carlo dynamics selects an \emph{edge} uniformly from the SBM edge list,
whereas the mean-field/ODE description is written in terms of a \emph{random neighbour} of a node.
For the equal-sized planted-partition SBM considered here, these lead to the same group-level mixing
statistics under an expected-degree approximation; the difference primarily rescales time and does not
change the eigenmode structure used below.

For the equal-sized planted-partition SBM with link probabilities $p_{\mathrm{in}}$ and $p_{\mathrm{out}}$,
an expected-degree approximation gives
\begin{equation}
\pi_{gg}=\frac{p_{\mathrm{in}}}{p_{\mathrm{in}}+(c-1)p_{\mathrm{out}}},
\qquad
\pi_{g\neq h}=\frac{p_{\mathrm{out}}}{p_{\mathrm{in}}+(c-1)p_{\mathrm{out}}}.
\end{equation}
The resulting spectrum contains the uniform eigenvalue $\lambda_1=1$ and a $(c-1)$-fold
degenerate modular (contrast) eigenvalue $\lambda_{\mathrm{mod}}=\pi_{gg}-\pi_{g\neq h}$.

\subsubsection*{Group-level Mean-field (ODE) Formulation}
\label{subsec:mf_ode_formulation}
To verify analytical boundaries and to isolate the role of initial conditions, we also integrate a deterministic group-level mean-field model defined directly in terms of the opinion fractions within each group. For each group $g$, let $f^{(g)}_{+}$, $f^{(g)}_{-}$, and $f^{(g)}_{0}$ denote the fractions of opinions $+1$, $-1$, and $0$, respectively. We work with the equivalent variables
\begin{equation}
m_g \equiv f^{(g)}_{+}-f^{(g)}_{-},
\qquad
s_g \equiv f^{(g)}_{+}+f^{(g)}_{-},
\qquad
f^{(g)}_{0}=1-s_g,
\label{eq:mf_defs_ms_model}
\end{equation}
where $m_g$ is the group mean opinion and $s_g$ is the group ``activity'' (fraction of non-neutral agents). Writing $\mathbf{m}=(m_1,\dots,m_c)^{T}$ and $\mathbf{s}=(s_1,\dots,s_c)^{T}$, and defining $\mathbf{M}=\Pi\mathbf{m}$ and $\mathbf{S}=\Pi\mathbf{s}$, the mean-field dynamics takes the form

\begin{align}
\dot{\mathbf m} &= (1-2p)\,\bigl(1-\tfrac{1}{2}\mathbf{s}\bigr)\odot \mathbf{M} \;-\; \tfrac{1}{2}\mathbf{S}\odot \mathbf{m}, \label{eq:ode_m_vec}\\
\dot{\mathbf s} &= \mathbf{S}\odot\bigl(1-\tfrac{3}{2}\mathbf{s}\bigr) \;+\; \tfrac{1}{2}(1-2p)\,\mathbf{m}\odot\mathbf{M}, \label{eq:ode_s_vec}
\end{align}
where $\odot$ denotes element-wise multiplication. The full derivation, including all intermediate steps of the gain-loss equations and the closure in terms of $(m_g, s_g)$, is given in the Appendix. The ODE observables are computed from $\mathbf{m}(t)$ as
\begin{equation}
O(t)=\left|\frac{1}{c}\sum_{g=1}^{c} m_g(t)\right|,
\qquad
O_{\mathrm{intra}}(t)=\frac{1}{c}\sum_{g=1}^{c}|m_g(t)|,
\qquad
\sigma(t)=\mathrm{std}\!\left(\{m_g(t)\}_{g=1}^{c}\right),
\label{eq:ode_obs_def}
\end{equation}
and stationary values in the phase maps are obtained by averaging $O(t)$ and $O_{\mathrm{intra}}(t)$ over a late-time window after transients.

\subsubsection*{ODE Initial Conditions}
\label{subsec:ode_ic_protocol}
Because the mean-field ODEs admit multiple attractors depending on symmetry sector, we explicitly control the initial condition of $\mathbf m(0)$ while fixing the initial activity near the active disordered value. Unless stated otherwise, seeded runs use
\begin{equation}
s_g(0)=\frac{2}{3}\quad \forall g,
\qquad
|m_g(0)|=\epsilon_m,
\label{eq:ode_seed_general}
\end{equation}
with a small seed magnitude $\epsilon_m$ and a prescribed sign pattern across groups.

For the two-group case ($c=2$) we use two symmetry-invariant classes:
(i) \emph{symmetric} seeds $m_1(0)=m_2(0)=\epsilon_m$, and
(ii) \emph{anti-symmetric} seeds $m_1(0)=+\epsilon_m$, $m_2(0)=-\epsilon_m$.
To emulate Monte Carlo-like disordered starts, we also use \emph{random} initial conditions generated by sampling group-level opinion fractions from a multinomial draw over $\{-1,0,+1\}$ and mapping them to $(m_g(0),s_g(0))$ via Equation~\eqref{eq:mf_defs_ms_model}; phase maps for this class are averaged over an ensemble of such random seeds.

For the multi-group case with even $c$, we consider four representative sign patterns:
(i) \emph{uniform} (all groups seeded with $+\epsilon_m$),
(ii) \emph{contrast} (half $+\epsilon_m$ and half $-\epsilon_m$),
(iii) \emph{random} (multinomial group-level initial fractions), and
(iv) \emph{almost-contrast} (one extra positively seeded group relative to the balanced contrast split).
These controlled initialisations are used to construct the ODE phase-space scans and trajectory panels in Section~\ref{sec:ch5_results}.

\section{Results and Discussions}
\label{sec:ch5_results}

\subsection*{Monte Carlo Simulations}
\label{subsec:mc_results}

\begin{figure}[htbp]
    \centering
    \includegraphics[width=0.98\linewidth]{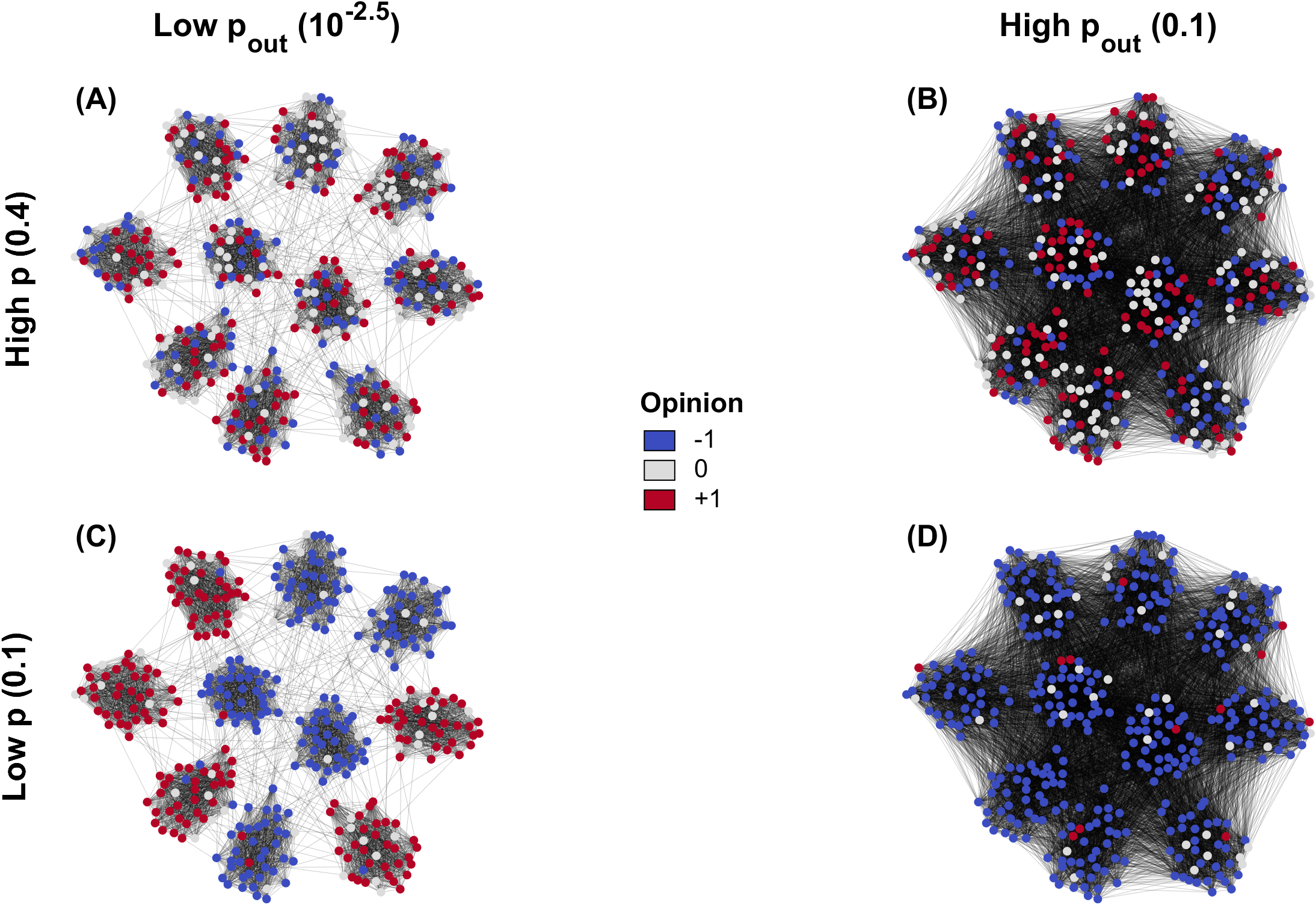}
    \includegraphics[width=0.98\linewidth]{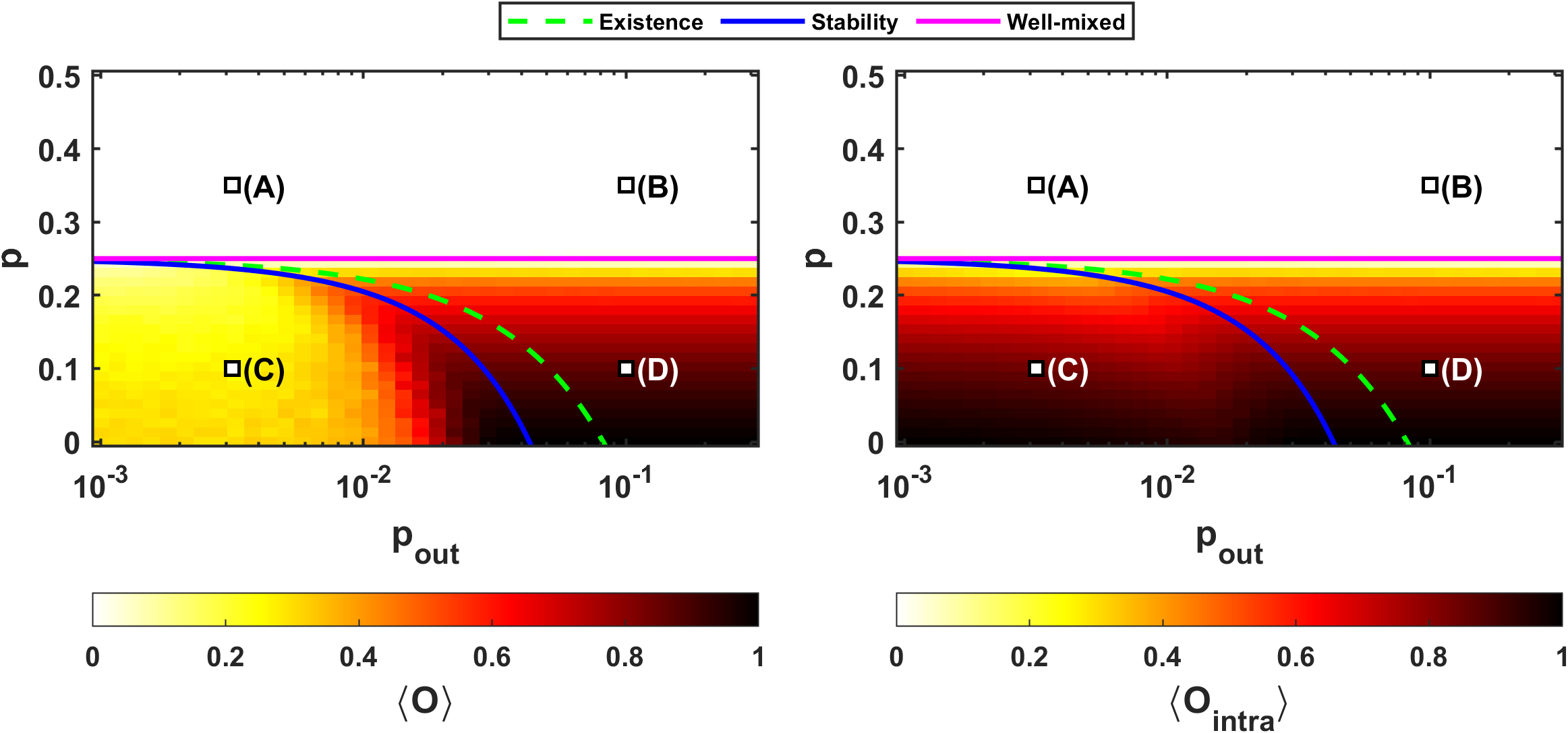}
    \caption{(A-D) Stationary opinion snapshots on modular SBM networks
    ($n{=}400$, $c{=}10$, $p_{\mathrm{in}}{=}0.99$) after $10^{7}$ BChS
    updates, with node colours encoding opinions $-1$ (blue), $0$ (grey),
    $+1$ (red). Left column: weak inter-group coupling
    ($p_{\mathrm{out}}{=}10^{-2.5}$); right column: strong coupling
    ($p_{\mathrm{out}}{=}0.1$). At high disagreement ($p{=}0.4$, top row)
    opinions remain disordered; at low disagreement ($p{=}0.1$, bottom row)
    weak coupling yields modular polarisation~(C) while strong coupling
    drives global consensus~(D). Lower panels: mean-field ODE phase diagrams
    of $\langle O\rangle$ and $\langle O_{\mathrm{intra}}\rangle$ in the
    $(p_{\mathrm{out}},\,p)$ plane for the same modular structure ($c=10$ and $p_{in}=.99$), obtained by ensemble-averaging over
    random initial conditions. Overlaid curves show the analytical
    contrast-mode existence (dashed green, Eq.~\eqref{eq:pexist_green}) and
    stability (solid blue, Eq.~\eqref{eq:pstable_blue}) boundaries, along with
    the well-mixed threshold $p{=}1/4$ (magenta). Squares mark the parameter
    values of panels~(A-D).}
    \label{fig1:snapshots}
\end{figure}

The regimes of collective behaviour are directly visible in the opinion snapshots (Figure~\ref{fig1:snapshots}). At high disagreement probability $p$ the dynamics generated by the BChS update rule (Equation~\eqref{eq:bchs_update}) produces disordered active states: opinions of both signs and the neutral state remain interspersed within each module, and both macroscopic order parameters are small. At low $p$, two distinct ordered outcomes are observed depending on inter-group coupling. For sufficiently weak $p_{\mathrm{out}}$, each module orders internally but the signs of the group-level mean opinions differ across modules, yielding a state with large intra-group order $O_{\mathrm{intra}}(t)$ (Equation~\eqref{eq:Ointra_def}) but small global order $O(t)$ (Equation~\eqref{eq:O_def}). This is modular polarisation (contrast ordering): strong local ordering with near-cancellation at the global level. Increasing $p_{\mathrm{out}}$ strengthens cross-module influence and promotes alignment of signs across groups, producing global consensus with both $O$ and $O_{\mathrm{intra}}$ large. The lower panels of Figure~\ref{fig1:snapshots} place these four regimes within the full $(p_{\mathrm{out}},p)$
 phase diagram obtained from the mean-field ODEs~\eqref{eq:mf_mdot_full}-\eqref{eq:mf_sdot_full}, with analytical existence~\eqref{eq:pexist_green} and stability~\eqref{eq:pstable_blue} boundaries overlaid.

\begin{figure}[htbp]
    \centering
    \includegraphics[width=0.98\linewidth]{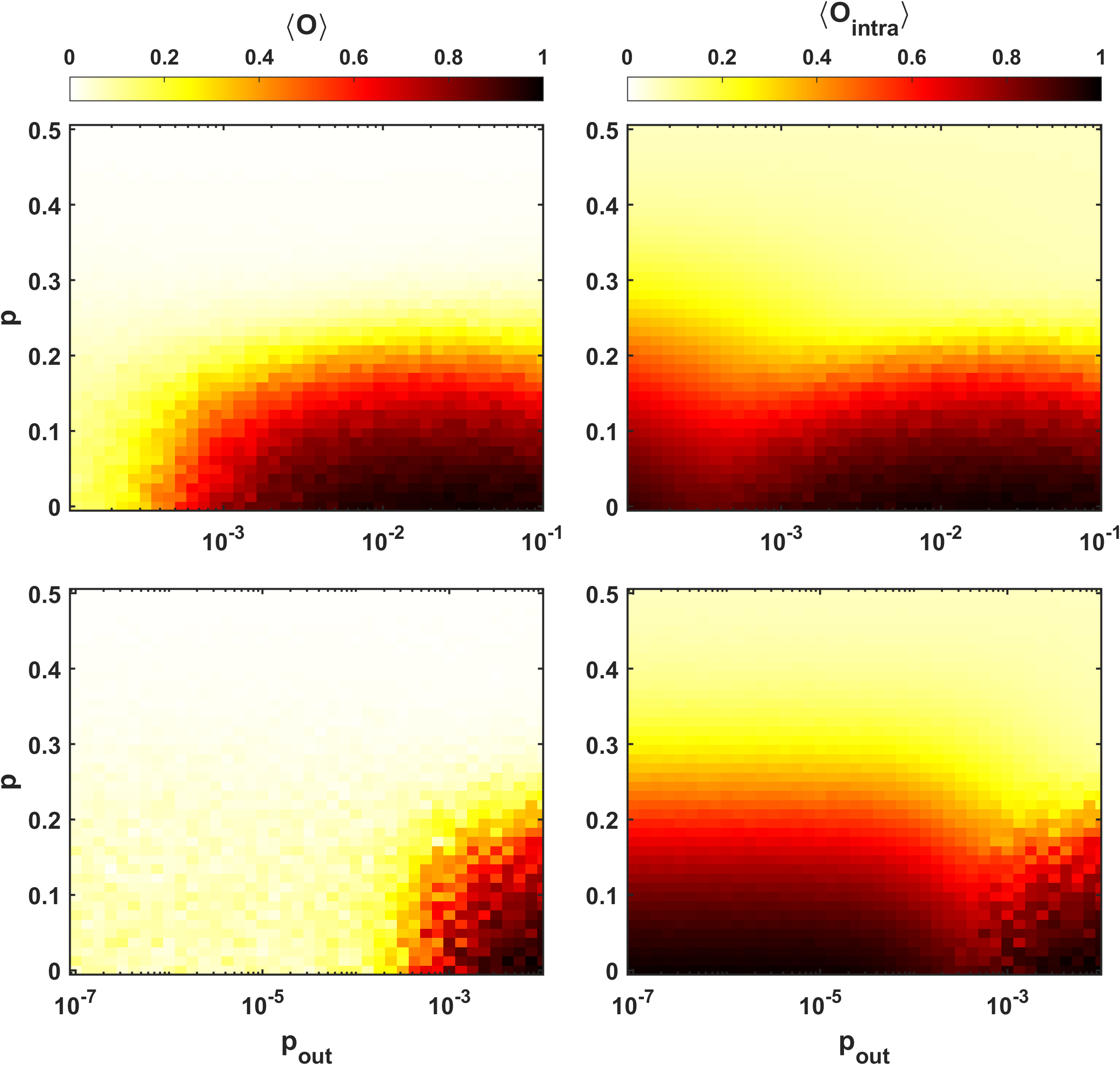}
    \caption{Global order parameter $O$ (left column) and intra-module order parameter $O_{\mathrm{intra}}$ (right column) in the $(p_{\mathrm{out}},p)$ plane for modular networks with $n = 10^{4}$ nodes partitioned into $c = 100$ equal groups. Networks are generated with a fixed $p_{\mathrm{in}}=0.9$ on a $42 \times 42$ grid of $(p_{\mathrm{out}},p)$, with $p_{\mathrm{out}}$ logarithmically spaced between $10^{-7}$ and $10^{-2}$ in the bottom row and between $10^{-3}$ and $10^{-1}$ in the top row, while $p$ is linearly spaced between $0$ and $0.5$ in all panels. For each parameter pair in the bottom row, observables are averaged over $nI = 6$ independent network realisations and over a stationary time window following transients of $T_{\mathrm{s}} = 2\times 10^{6}$ steps and a measurement period of $T_{\mathrm{m}} = 2\times 10^{6}$ steps; the top row shows the same quantities for a narrower $p_{\mathrm{out}}$ range with $nI = 50$ realisations per point, providing a higher-precision estimate of the phase-diagram structure. Colour encodes the averaged values of the corresponding order parameter (see colour bars), illustrating how increasing inter-group connectivity $p_{\mathrm{out}}$ and disagreement probability $p$ drive the system from disordered to globally and modularly ordered regimes.}
    \label{fig2:phase_pout_p}
\end{figure}
These trends, confirmed by the ODE phase diagram for $c=10$ in Figure~\ref{fig1:snapshots},  persist at larger system size and number of modules. Figure~\ref{fig2:phase_pout_p} summarises Monte Carlo phase diagrams in the $(p_{\mathrm{out}},p)$ plane for $c=100$ (fixed $p_{\mathrm{in}}=0.9$). At large $p$, $\langle O\rangle\approx 0$ and $\langle O_{\mathrm{intra}}\rangle\approx 0$ across all $p_{\mathrm{out}}$, consistent with a stable disordered active regime. Decreasing $p$ first produces a broad low-$p_{\mathrm{out}}$ region where $\langle O_{\mathrm{intra}}\rangle$ is large but $\langle O\rangle$ remains small, i.e.\ modular polarisation. Only when $p_{\mathrm{out}}$ exceeds a $p$-dependent threshold does $\langle O\rangle$ increase sharply, indicating the onset of global consensus.

The separation between within-group ordering and global consensus is naturally interpreted in a mode-based mean-field description, where the SBM enters only through the neighbour-mixing matrix $\Pi$ and its eigenmodes.

\subsection*{Mean-field Approach}
\label{subsec:mf_linear_baseline}

A compact mean-field formulation is obtained by tracking, for each group $g=1,\dots,c$, the opinion fractions $f^{(g)}_{+}$, $f^{(g)}_{-}$, and $f^{(g)}_{0}$, with $f^{(g)}_{+}+f^{(g)}_{-}+f^{(g)}_{0}=1$. It is convenient to work with the group-level mean opinion and activity,
\begin{equation}
m_g \equiv f^{(g)}_{+}-f^{(g)}_{-},
\qquad
s_g \equiv f^{(g)}_{+}+f^{(g)}_{-},
\qquad
f^{(g)}_{0}=1-s_g,
\label{eq:mf_defs_ms}
\end{equation}
which correspond, respectively, to the signed quantity underlying the order parameters in Equations~\eqref{eq:O_def} and \eqref{eq:Ointra_def}.

Modularity enters through the neighbour-mixing matrix $\Pi=[\pi_{gh}]$ (Section~\ref{subsec:sbm_mapping}), where $\pi_{gh}$ is the probability that a random neighbour of a node in group $g$ lies in group $h$. Defining the neighbour-averaged fields
\begin{equation}
M_g = \sum_{h=1}^{c}\pi_{gh} m_h,
\qquad
S_g = \sum_{h=1}^{c}\pi_{gh} s_h,
\label{eq:MS_fields}
\end{equation}
the mean-field ODEs for BChS dynamics can be written in closed form as
\begin{align}
\dot m_g &= (1-2p)\left(1-\frac{s_g}{2}\right) M_g \;-\; \frac{S_g}{2}\,m_g,
\label{eq:mf_mdot_full}\\
\dot s_g &= S_g\left(1-\frac{3}{2}s_g\right) \;+\; \frac{1-2p}{2}\,m_g M_g.
\label{eq:mf_sdot_full}
\end{align}
These equations provide a deterministic baseline that isolates how mixing eigenmodes control the onset of ordering.

\paragraph*{Disordered Fixed Point.}
A disordered state has $m_g=0$ for all $g$. Substituting $m_g=0$ into Equation~\eqref{eq:mf_sdot_full} yields the group-symmetric active fixed point
\begin{equation}
m_g^{*}=0\;\;\forall g,
\qquad
s_g^{*}=\frac{2}{3}\;\;\forall g,
\label{eq:mf_disordered_fp}
\end{equation}
equivalently $f^{(g)}_{+}=f^{(g)}_{-}=f^{(g)}_{0}=1/3$ in every group.

\paragraph*{Linear Stability.}
Near the disordered active state, the coupling of $\dot s_g$ to magnetisation enters only through the quadratic term $m_gM_g$ in Equation~\eqref{eq:mf_sdot_full}. Therefore, to linear order one may set $s_g\simeq S_g\simeq 2/3$ in Equation~\eqref{eq:mf_mdot_full}, yielding
\begin{equation}
\dot{\mathbf m}
=
\left[
\frac{2}{3}(1-2p)\,\Pi
-\frac{1}{3}I
\right]\mathbf m,
\label{eq:mf_linear_m}
\end{equation}
where $\mathbf m=(m_1,\dots,m_c)^{T}$. If $\Pi\mathbf v_\alpha=\lambda_\alpha\mathbf v_\alpha$, then the growth rate of mode $\alpha$ is
\begin{equation}
\Lambda_\alpha=\frac{2}{3}(1-2p)\lambda_\alpha-\frac{1}{3},
\label{eq:mf_growth_rates}
\end{equation}
and $\Lambda_\alpha=0$ gives the critical line (Equation~\eqref{eq:mf_pc_lambda})
\begin{equation}
p_c^{(\alpha)}=\frac{1}{2}\left(1-\frac{1}{2\lambda_\alpha}\right).
\label{eq:mf_pc_lambda}
\end{equation}
The uniform eigenmode $\lambda_1=1$ recovers the well-mixed threshold $p_c=1/4$. The remaining eigenmodes correspond to group-contrast patterns. For the equal-sized planted-partition SBM, $\Pi$ has a $(c-1)$-fold degenerate contrast eigenvalue $\lambda_{\mathrm{mod}}=\pi_{gg}-\pi_{g\neq h}$, so Equation~\eqref{eq:mf_pc_lambda} predicts a distinct contrast-mode instability associated with modular ordering. In the planted-partition parameterisation used for the ODE phase diagrams,
\begin{equation}
\Pi_{gg}=a,
\qquad
\Pi_{g\neq h}=\frac{1-a}{c-1},
\qquad
\lambda_{\mathrm{mod}}=\frac{ac-1}{c-1}.
\label{eq:Pi_a_def}
\end{equation}

Equations~\eqref{eq:mf_mdot_full}--\eqref{eq:mf_pc_lambda} should be read as a baseline: they determine when the disordered active state loses stability along a given eigenmode, but do not by themselves classify nonlinear attractors or basin structure. The next subsections use $c=2$ to make the symmetry manifolds explicit and then use a two-supercluster reduction to obtain analytical existence and stability boundaries for the contrast-ordered branch.

\subsection*{Special Case of Two Groups}

The two-group limit is distinguished by two symmetry-invariant manifolds, corresponding to the uniform and contrast modes. For $c=2$ the mixing matrix is
\begin{equation}
\Pi=
\begin{pmatrix}
a & 1-a\\
1-a & a
\end{pmatrix},
\qquad
\lambda \equiv \lambda_{\mathrm{mod}} = 2a-1,
\label{eq:Pi_c2}
\end{equation}
with eigenvectors proportional to $(1,1)$ (uniform) and $(1,-1)$ (contrast). The symmetric manifold $m_1=m_2$ and $s_1=s_2$ reduces the dynamics exactly to the one-group mean-field and yields the usual consensus transition at $p=1/4$. The antisymmetric manifold $m_2=-m_1$ with equal activities $s_1=s_2\equiv s$ corresponds to contrast ordering with vanishing global mean.

On the antisymmetric manifold, $M_1=\lambda m_1$ and $M_2=\lambda m_2$, while $S_1=S_2=s$. Equation~\eqref{eq:mf_mdot_full} becomes
\begin{equation}
\dot m_1=\left[(1-2p)\lambda\left(1-\frac{s}{2}\right)-\frac{s}{2}\right]m_1.
\label{eq:c2_antisym_mdot}
\end{equation}
Defining the effective control parameter
\begin{equation}
q \equiv (1-2p)\lambda,
\label{eq:q_def}
\end{equation}
the non-trivial antisymmetric branch follows from the fixed-point conditions of Equations~\eqref{eq:mf_mdot_full}--\eqref{eq:mf_sdot_full}:
\begin{equation}
s_-=\frac{2q}{1+q},
\qquad
m_1^2=\frac{4(2q-1)}{(1+q)^2}.
\label{eq:c2_antisym_branch}
\end{equation}
Hence the antisymmetric (contrast-ordered) branch exists only when
\begin{equation}
q>\frac{1}{2}
\quad\Longleftrightarrow\quad
(1-2p)(2a-1)>\frac{1}{2}.
\label{eq:c2_existence_q}
\end{equation}
This immediately explains the bipartite-contrarian sector: for $p>1/2$ one has $(1-2p)<0$, and for $a<1/2$ one has $\lambda=2a-1<0$, so $q$ can still be positive and exceed $1/2$, supporting stable opposite-sign group mean opinions even at high contrarian probability.

\begin{figure}[htbp]
\centering
\begin{overpic}[width=0.98\linewidth]{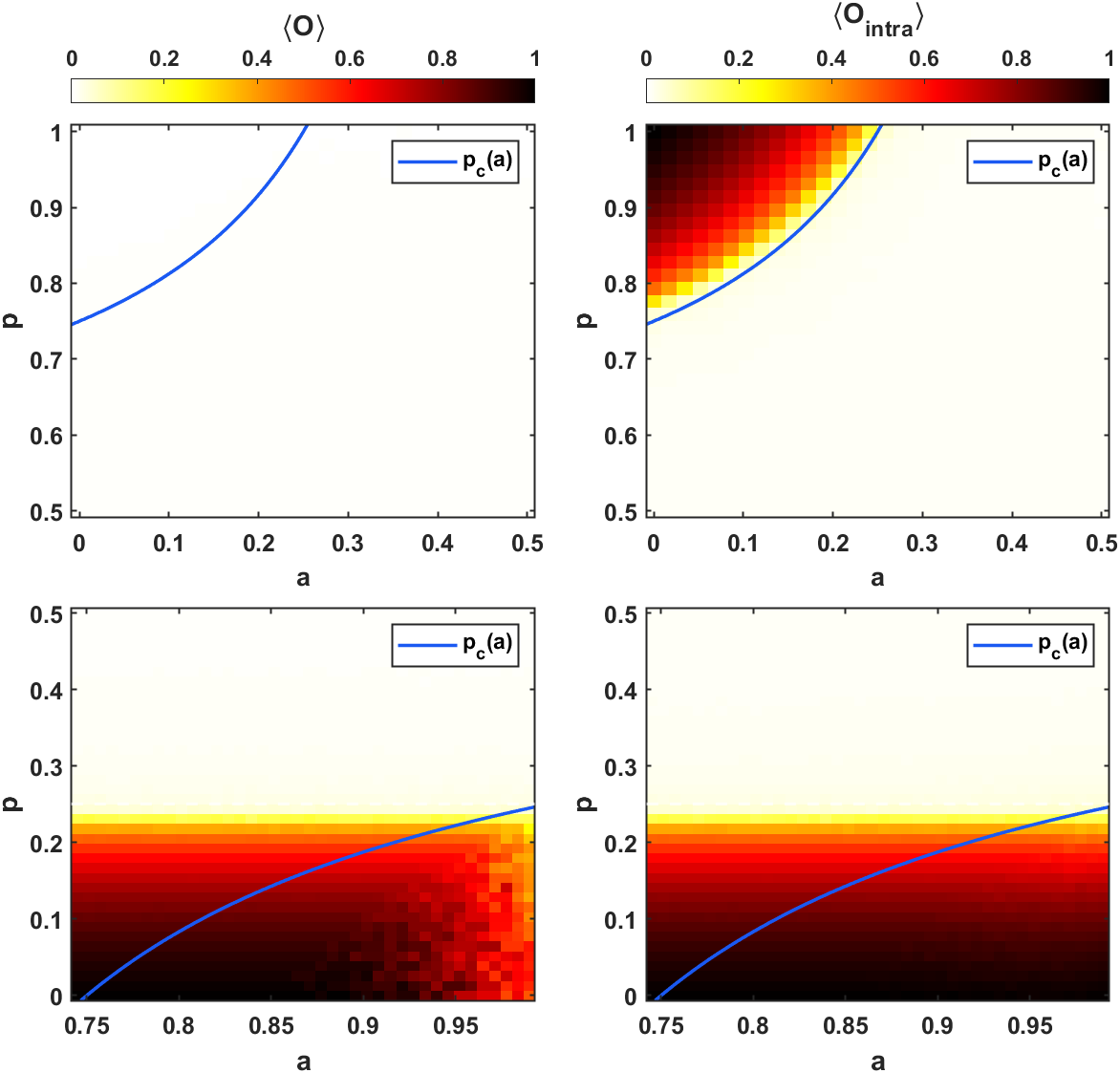}
  \put(1,93){\textbf{A.}}
  \put(51,93){\textbf{B.}}
  \put(1,44){\textbf{C.}}
  \put(51,44){\textbf{D.}}
\end{overpic}
\caption{\textbf{Two-group Monte Carlo phase-space scans highlighting the bipartite and consensus-ordering regimes.}
BChS dynamics on stochastic block networks with $c=2$ groups, $n=10^{4}$ agents, and fixed average degree $\langle k\rangle=20$, shown in the $(a,p)$ plane, where $a$ is the self-mixing probability and $p$ is the contrarian probability. (A,B) Scan on a $30\times 30$ grid over $a\in[0,0.5]$ and $p\in[0.5,1]$, corresponding to the high-$p$, low-$a$ sector that contains the bipartite-contrarian ordered regime; the stationary averages $\langle O\rangle$ and $\langle O_{\mathrm{intra}}\rangle$ are shown in (A) and (B), respectively. (C,D) Refined scan on a $40\times 40$ grid over $a\in[0.745,0.99]$ and $p\in[0,0.5]$, corresponding to the high-$a$, low-$p$ sector associated with the conventional consensus-ordering regime; $\langle O\rangle$ and $\langle O_{\mathrm{intra}}\rangle$ are shown in (C) and (D), respectively. For each grid point, observables are averaged over independent realisations (\texttt{NumIter}=30 in (A,B) and \texttt{NumIter}=40 in (C,D)) after a transient of $T_{\mathrm{s}}=2\times 10^{6}$ update steps and over a measurement window of length $T_{\mathrm{m}}=2\times 10^{6}$ steps. Colour encodes the corresponding stationary average (colour bars). The blue curve denotes the analytical contrast-mode instability boundary $p_c(a)$.}
\label{fig:ch5_fig3_2group}
\end{figure}

Figure~\ref{fig:ch5_fig3_2group} shows that the two ordered regimes correspond to distinct symmetry sectors. In panels C--D (high $a$, low $p$), uniform-mode ordering produces consensus with large $\langle O\rangle$ and large $\langle O_{\mathrm{intra}}\rangle$. In panels A--B (low $a$, high $p$), contrast ordering produces large $\langle O_{\mathrm{intra}}\rangle$ but small $\langle O\rangle$, consistent with the antisymmetric branch \eqref{eq:c2_antisym_branch}. The analytical curve provides the linear contrast-mode onset (Equation~\eqref{eq:mf_pc_lambda} with $\lambda=2a-1$), which is the appropriate baseline for locating where the disordered active state becomes unstable to contrast perturbations.

\begin{figure}[htbp]
\centering
\begin{overpic}[width=0.98\linewidth]{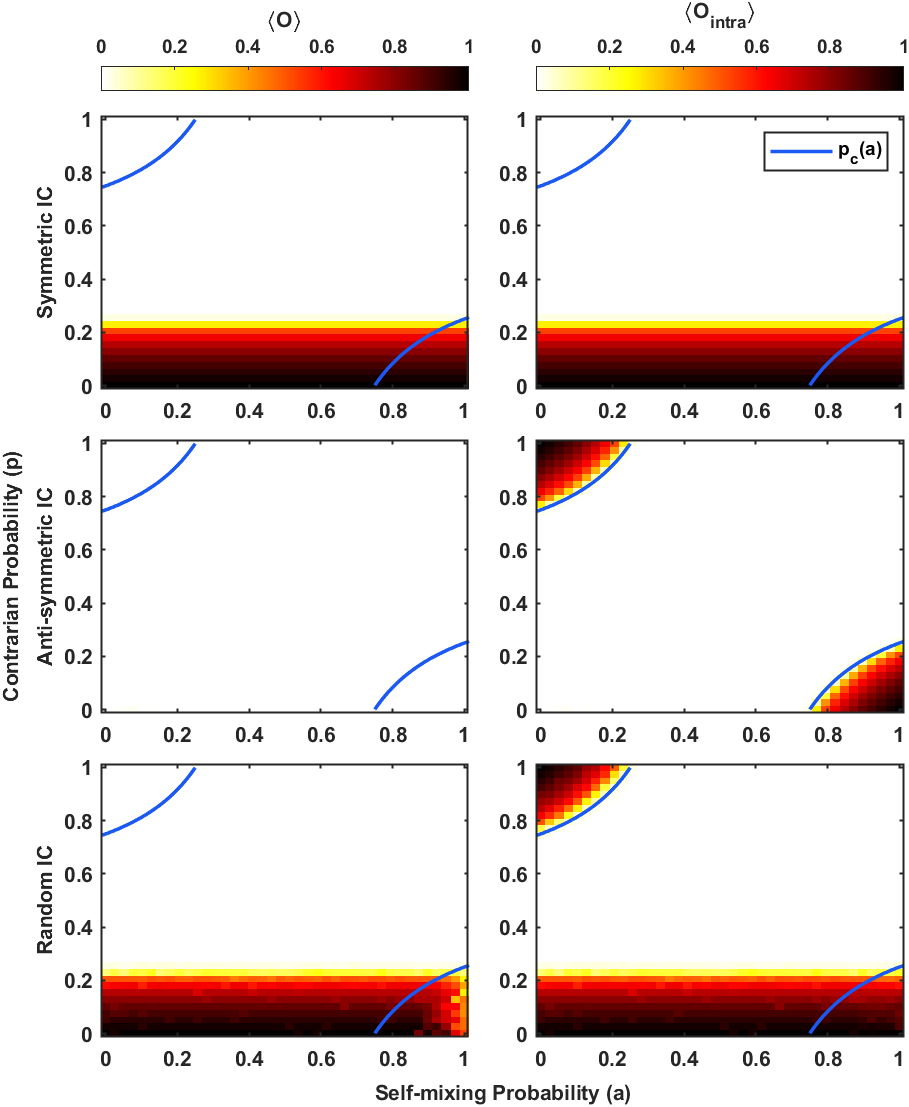}
  \put(4,91){\textbf{A.}}
  \put(44,91){\textbf{B.}}
  \put(4,62){\textbf{C.}}
  \put(44,62){\textbf{D.}}
  \put(4,32){\textbf{E.}}
  \put(44,32){\textbf{F.}}
\end{overpic}
\caption{\textbf{Two-group ODE phase-space scans for different initial-conditions.}
Stationary phase-space maps of the two-group ($c=2$) BChS dynamics in the $(a,p)$ plane. The left column shows $\langle O\rangle$ and the right column shows $\langle O_{\mathrm{intra}}\rangle$. (A,B) Symmetric initial condition, $m_1(0)=m_2(0)=\epsilon_m$. (C,D) Anti-symmetric initial condition, $m_1(0)=+\epsilon_m$, $m_2(0)=-\epsilon_m$. (E,F) Random initial condition from MC-like multinomial opinion fractions. In all cases, $\epsilon_m=5\times10^{-2}$ and $s_1(0)=s_2(0)=2/3$ for seeded runs. All panels use a $40\times40$ grid over $a,p\in[0,1]$. Colour encodes the stationary average (colour bars in the top row). The blue curve denotes the analytical contrast-mode instability boundary $p_c(a)$.}
\label{fig:ch5_fig4_bipartite_ode}
\end{figure}

The ODE phase spaces in Figure~\ref{fig:ch5_fig4_bipartite_ode} make the basin structure explicit. Symmetric seeding (A--B) excites only the uniform mode and therefore selects the consensus branch whenever it exists. Anti-symmetric seeding (C--D) isolates the contrast manifold and tests the existence of the antisymmetric branch \eqref{eq:c2_antisym_branch}. Random (MC-like) initial conditions (E--F) contain both uniform and contrast components, and therefore probe competition between consensus and contrast attractors. This explains why the same $(a,p)$ region can exhibit different stationary outcomes in deterministic ODE scans depending on initial-condition class, while Monte Carlo simulations naturally include finite-size symmetry breaking and noise-induced exploration of basins.

\begin{figure}[t]
\centering
\begin{overpic}[width=0.98\linewidth]{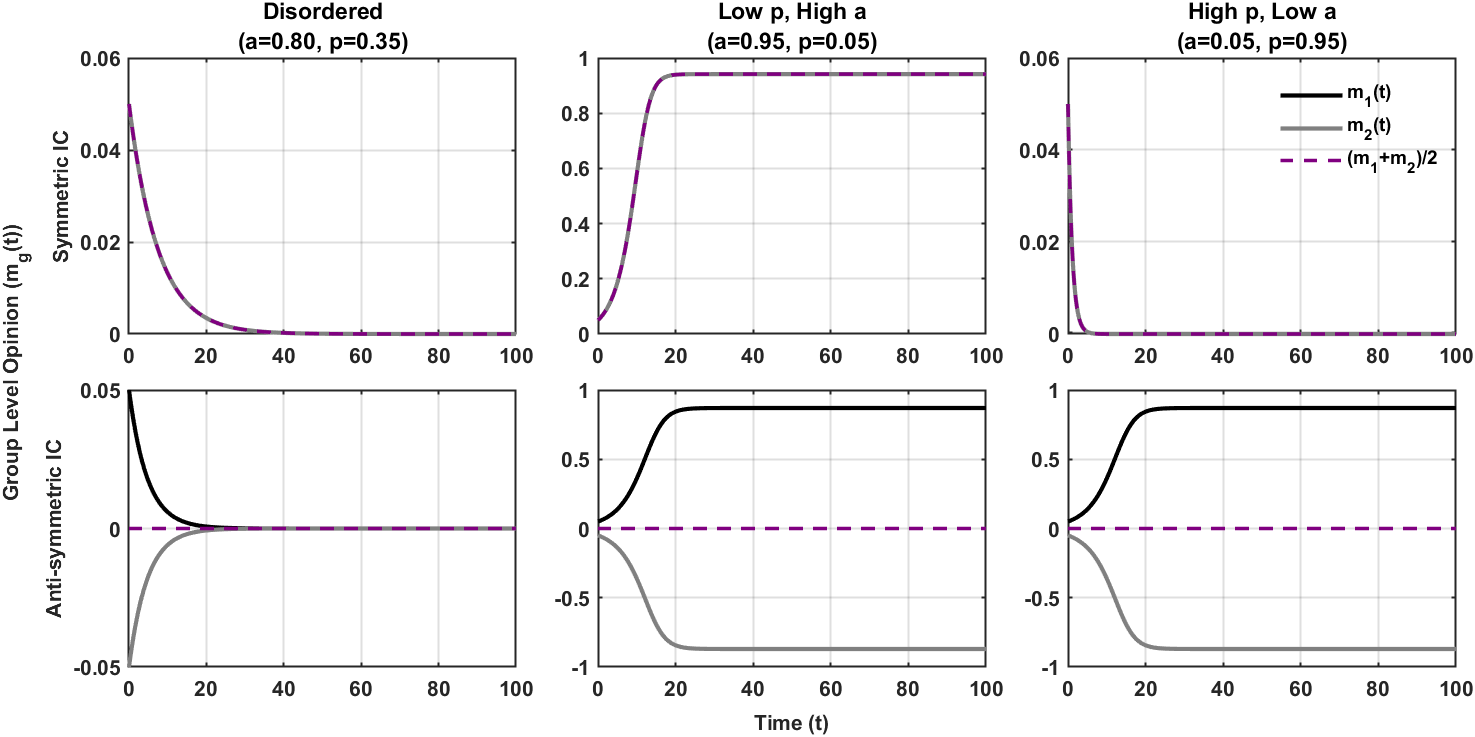}
  \put(2,48){\textbf{A.}}
  \put(36,48){\textbf{B.}}
  \put(68,48){\textbf{C.}}
  \put(2,23){\textbf{D.}}
  \put(36,23){\textbf{E.}}
  \put(68,23){\textbf{F.}}
\end{overpic}
\caption{\textbf{Two-group ODE trajectories in representative regimes.}
Time evolution of the two-group ($c=2$) mean-field BChS dynamics, showing $m_1(t)$, $m_2(t)$, and the mean group opinion $(m_1+m_2)/2$. Columns show the disordered regime (A,D), the low-$p$/high-$a$ consensus-ordering regime (B,E), and the high-$p$/low-$a$ bipartite-contrarian regime (C,F), with $(a,p)=(0.80,0.35)$, $(0.95,0.05)$, and $(0.05,0.95)$, respectively. The top row (A--C) uses a symmetric initial condition, $m_1(0)=m_2(0)=\epsilon_m$, and the bottom row (D--F) uses an anti-symmetric initial condition, $m_1(0)=+\epsilon_m$, $m_2(0)=-\epsilon_m$, with $\epsilon_m=5\times 10^{-2}$ and $s_1(0)=s_2(0)=2/3$. The trajectories show relaxation to the disordered state, global consensus, and opposite-sign group ordering in the three regimes.}
\label{fig:ch5_fig5_bipartite_traj}
\end{figure}

The representative trajectories in Figure~\ref{fig:ch5_fig5_bipartite_traj} provide the dynamical interpretation. In the disordered regime, both group-level mean opinions decay to zero. In the consensus regime, $m_1(t)$ and $m_2(t)$ converge to the same sign, giving large $O$ and large $O_{\mathrm{intra}}$. In the bipartite-contrarian regime, the two groups converge to equal-magnitude opposite-sign mean opinions, consistent with the antisymmetric fixed point \eqref{eq:c2_antisym_branch}; the mean $(m_1+m_2)/2$ vanishes while $|m_1|$ and $|m_2|$ remain large, giving large $O_{\mathrm{intra}}$ but small $O$.

\subsection*{Two-supercluster Reduction Approach}
\label{subsec:green_blue}

For $c\gg 1$, the contrast-ordered regime can be organised by a two-supercluster reduction in which groups split into two blocks with opposite-sign mean opinions but comparable activity. In this setting it is useful to express the contrast eigenvalue explicitly in terms of $(a,c)$:
\begin{equation}
\lambda(c)=\lambda_{\mathrm{mod}}=\frac{ac-1}{c-1}.
\label{eq:lambda_c}
\end{equation}
Define the effective control parameter
\begin{equation}
q \equiv (1-2p)\lambda(c).
\label{eq:q_def_c}
\end{equation}
On the balanced contrast manifold (equal numbers of $+m$ and $-m$ groups), the non-trivial branch obeys the same algebraic structure as the $c=2$ antisymmetric solution:
\begin{equation}
s=\frac{2q}{1+q},
\qquad
m^2=\frac{4(2q-1)}{(1+q)^2},
\label{eq:branch_ms_generalc}
\end{equation}
hence it exists only when $q>1/2$.

\paragraph*{Existence Boundary.}
Using the disordered activity $s^{*}=2/3$ to locate the onset of contrast ordering gives the existence line
\begin{equation}
p_{\mathrm{exist}}(a,c)
=
\frac{1}{2}\left(1-\frac{1}{2\lambda(c)}\right),
\label{eq:pexist_green}
\end{equation}
which is the contrast-mode instability line (Equation~\eqref{eq:mf_pc_lambda}) evaluated on $\lambda=\lambda(c)$.

\paragraph*{Stability Boundary.}
Existence does not guarantee stability of the balanced contrast branch against perturbations that inject a uniform (consensus) component. The loss of stability occurs when a coupled uniform-magnetisation/contrast-activity perturbation becomes marginal, which yields a quadratic condition for the critical $q$ as a function of $\lambda$:
\begin{equation}
(1+\lambda)\,q^{2} + (2\lambda^{2}+\lambda-2)\,q - \lambda^{2}=0.
\label{eq:qcrit_quadratic}
\end{equation}
Let $q_c(\lambda)$ denote the positive root of Equation~\eqref{eq:qcrit_quadratic}. The corresponding stability boundary in the $(a,p)$ plane is
\begin{equation}
p_{\mathrm{stable}}(a,c)
=
\frac{1}{2}\left(1-\frac{q_c(\lambda(c))}{\lambda(c)}\right).
\label{eq:pstable_blue}
\end{equation}

\begin{figure}[htbp]
\centering
\begin{overpic}[width=0.98\linewidth]{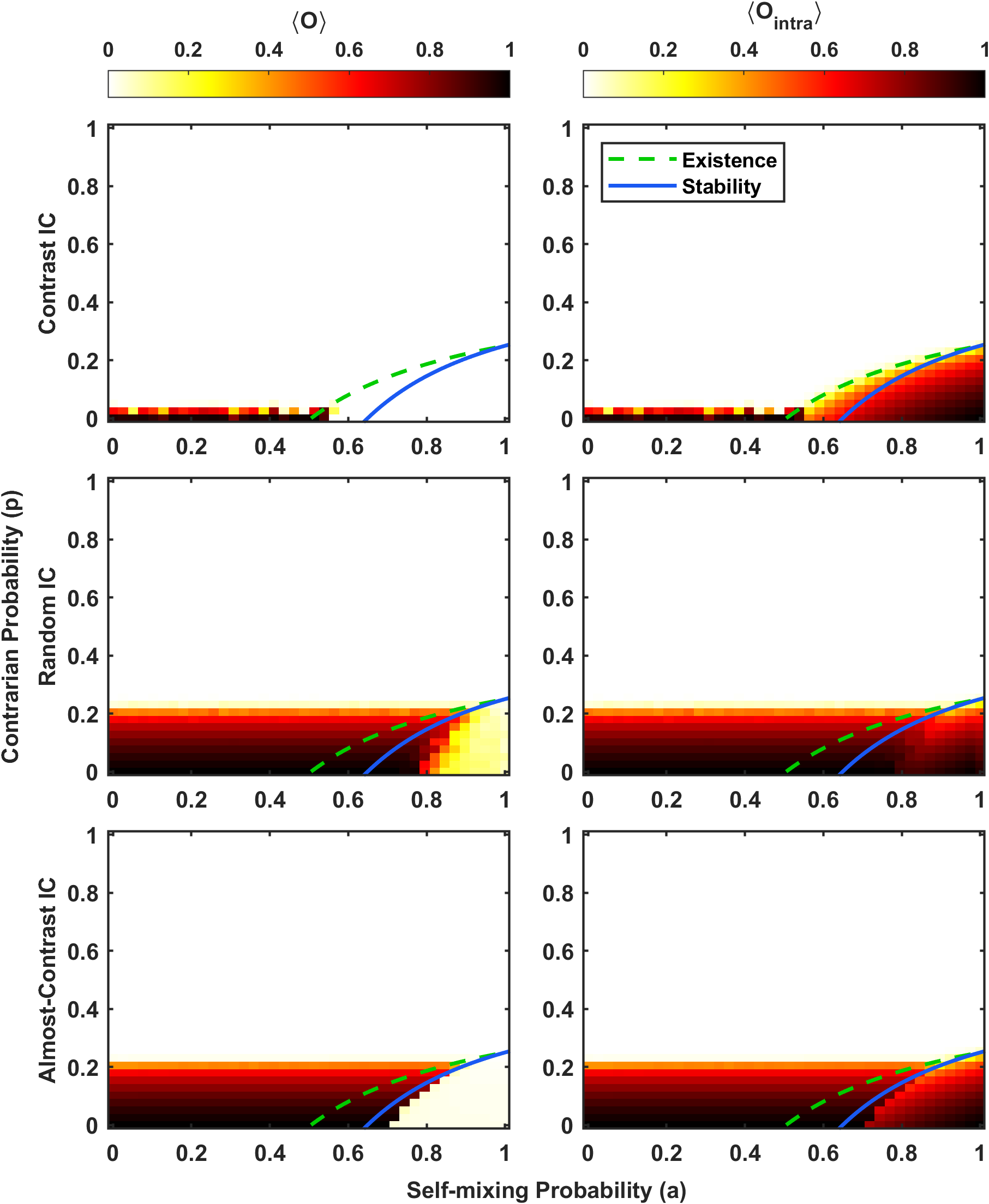}
  \put(4,91){\textbf{A.}}
  \put(44,91){\textbf{B.}}
  \put(4,62){\textbf{C.}}
  \put(44,62){\textbf{D.}}
  \put(4,32){\textbf{E.}}
  \put(44,32){\textbf{F.}}
\end{overpic}
\caption{\textbf{ODE phase diagrams for the multi-group case for different initial conditions.}
Phase-space maps of $\langle O\rangle$ (left column) and $\langle O_{\mathrm{intra}}\rangle$ (right column) on the full $(a,p)$ plane for the $c=100$ case. (A,B) Contrast initial condition (half the groups seeded with $+\epsilon_m$, half with $-\epsilon_m$). (C,D) Random initial condition from MC-like multinomial sampling. (E,F) Almost-contrast initial condition (one extra positively seeded group). In all panels, $s_g(0)=2/3$ and $\epsilon_m=0.1$ for seeded runs. Colour encodes the stationary value of the corresponding order parameter. The dashed green and solid blue curves denote the analytical existence and stability boundaries, respectively, for the contrast branch.}
\label{fig:ch5_fig6_multigroup_ode_phase}
\end{figure}

Figure~\ref{fig:ch5_fig6_multigroup_ode_phase} compares these analytical curves to the ODE phase maps for $c=100$. The dashed green line (Equation~\eqref{eq:pexist_green}) marks where the contrast-ordered branch becomes available. The solid blue line (Equation~\eqref{eq:pstable_blue}) marks where that branch is stable against drift toward consensus. The region between these curves is therefore expected to show initial-condition dependence: a contrast-seeded state can remain modularly polarised, while initial conditions containing a sufficient uniform component can evolve toward consensus even though the contrast branch exists.

\subsection*{ODE Simulations}
\label{subsec:ode_verif}

Because the ODE dynamics is deterministic, the observed stationary state depends on which symmetry sector is initially excited. In particular, a perfectly balanced contrast initial condition lies close to the $\sum_g m_g=0$ manifold and suppresses the uniform mode, making it effective for testing existence of the contrast branch. By contrast, random (MC-like) initial conditions generally contain both uniform and contrast components, and therefore probe stability against consensus drift. This is the appropriate deterministic analogue of Monte Carlo simulations, where finite-size fluctuations continuously seed weak uniform components.

\begin{figure}[htbp]
    \centering
    \begin{overpic}[width=0.98\linewidth]{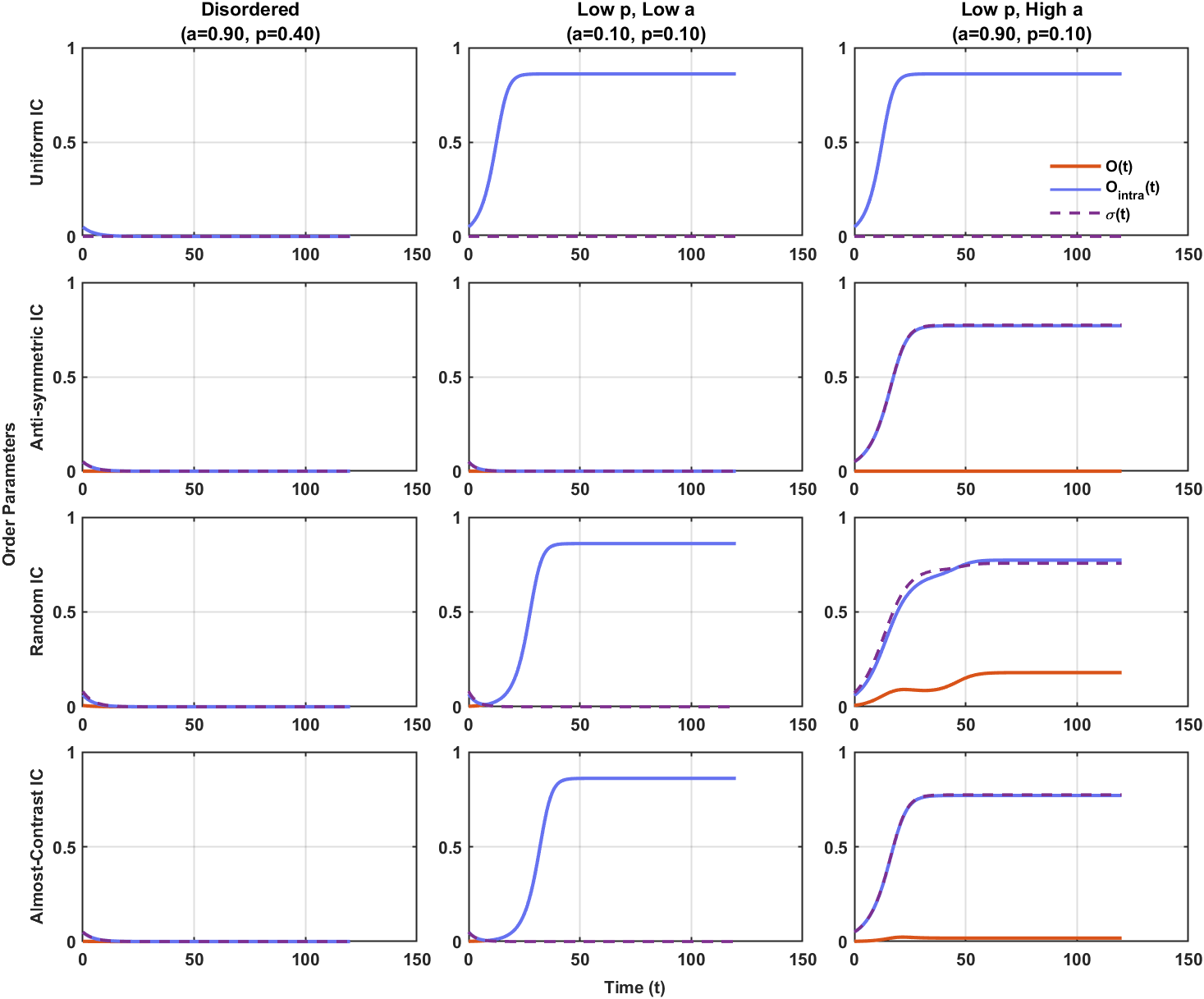}
        \put(3,79.2){\textbf{A.}}
        \put(35.3,79.2){\textbf{B.}}
        \put(67.5,79.2){\textbf{C.}}

        \put(3,59.5){\textbf{D.}}
        \put(35.3,59.5){\textbf{E.}}
        \put(67.5,59.5){\textbf{F.}}

        \put(3,40){\textbf{G.}}
        \put(35.3,40){\textbf{H.}}
        \put(67.5,40){\textbf{I.}}

        \put(3,20.5){\textbf{J.}}
        \put(35.3,20.5){\textbf{K.}}
        \put(67.5,20.5){\textbf{L.}}
    \end{overpic}
    \caption{\textbf{ODE trajectories of multi-group order parameters for four initial conditions.}
Panels show the mean-field ODE dynamics for $c=100$ groups in three parameter regimes: disordered (left column, $a=0.90$, $p=0.40$), low-$p$ low-$a$ (middle column, $a=0.10$, $p=0.10$), and low-$p$ high-$a$ (right column, $a=0.90$, $p=0.10$). Rows correspond to the four initial conditions: uniform (A--C), contrast (D--F), random MC-like initial opinions (G--I), and almost-contrast (J--L). In each panel, the time series of the global order parameter $O(t)$, intra-group order parameter $O_{\mathrm{intra}}(t)$, and the spread $\sigma(t)=\mathrm{std}(m_g)$ are shown. All trajectories use the same initial activity $s_g(0)=2/3$ and a small initial opinion seed magnitude $\epsilon_m=0.05$ (with the sign pattern set by the chosen initial condition). The figure highlights the expected decay to disorder at high $p$, global ordering at low $p$ and low $a$, and sustained modular ordering at low $p$ and high $a$, with the random and almost-contrast initial conditions showing intermediate transients before converging to the corresponding attractors.}
    \label{fig:Fig5_7}
\end{figure}

Figure~\ref{fig:Fig5_7} confirms the analytical organisation by symmetry modes in the multi-group setting. In the disordered regime (left column), $O(t)$ and $O_{\mathrm{intra}}(t)$ decay for all initial conditions, consistent with stability of the disordered active state. In the consensus regime (middle column), trajectories converge to a uniform-sign state: $O$ and $O_{\mathrm{intra}}(t)$ become large while the spread $\sigma(t)$ collapses, indicating loss of contrast structure. In the modular regime (right column), contrast-type initial conditions maintain large $O_{\mathrm{intra}}(t)$ and non-zero spread $\sigma(t)$ while $O(t)$ remains small, consistent with stable contrast ordering. In the parameter region where the contrast branch exists but is not stable to consensus drift (between the dashed green and solid blue boundaries in Figure~\ref{fig:ch5_fig6_multigroup_ode_phase}), the long-time outcome depends on whether the initial condition contains a sufficiently strong uniform component, as expected from the stability criterion encoded in Equations~\eqref{eq:qcrit_quadratic} and \eqref{eq:pstable_blue}.

\section{Summary and Conclusions}
\label{sec:ch5_conclusion}

In this work, we investigated the Biswas--Chatterjee--Sen (BChS) kinetic-exchange dynamics on modular interaction networks generated by stochastic block models. Monte Carlo simulations showed that modularity separates two distinct ordering scales. For sufficiently weak inter-group coupling, the system can sustain internally ordered modules whose mean opinions cancel at the global level. In contrast, stronger inter-group coupling promotes alignment across modules and drives the system toward global consensus. These regimes persist at large system sizes and organise naturally in the $(p_{\mathrm{out}},p)$ phase space.

A central feature of the phase structure is the emergence of a bipartite-contrarian ordered phase in the two-group limit. In the low-$a$, high-$p$ sector--where
between-group interactions dominate and most exchanges are repulsive--the system does not relax to disorder. Instead, it converges to a stable antisymmetric state with equal-magnitude, opposite-sign group mean opinions. This phase is associated with the contrast eigenmode of the mixing matrix and shows that sufficiently segregated structure can stabilise polarised order even when the microscopic rule favours disagreement. In this regime, disagreement does not destroy order; rather, it supports structured polarisation across groups.

To interpret these outcomes, we formulated a group-level mean-field description in which modular structure enters through the neighbour-mixing matrix $\Pi$. Linearisation about the disordered active state showed that the onset of ordering is controlled by the eigenmodes of $\Pi$: the uniform mode reproduces the well-mixed threshold, while the contrast eigenspace provides a distinct instability associated with modular ordering. In the two-group case, this mode-based picture explains the existence of the antisymmetric branch in terms of the effective control parameter $q=(1-2p)\lambda$.

Characterising stability beyond linear onset requires analysis of the modular branch itself. Using a two-supercluster reduction, we derived analytical existence and stability boundaries for the balanced contrast state. The stability limit is obtained from a reduced Jacobian that captures coupling between the uniform magnetisation mode and contrast-sector activity fluctuations. Comparison with numerical ODE phase maps confirms that modular ordering is not determined solely by linear instability: between the existence and stability boundaries, the long-time outcome depends on the symmetry sector excited by the initial condition.

Taken together, these results quantify how modular mixing reshapes the route to collective order. The same microscopic interaction rule can generate global consensus, sustained modular polarisation, or bipartite-contrarian ordering depending solely on the spectral structure of the interaction network.

\section*{Appendices}

\setcounter{section}{0}
\renewcommand{\thesection}{A\arabic{section}}
\setcounter{figure}{0} \setcounter{table}{0} \renewcommand{\thefigure}{A\arabic{figure}} \renewcommand{\thetable}{A\arabic{table}}

%This appendix presents the analytical derivation used to interpret the modular phase structure of the BChS dynamics on stochastic block networks. The development proceeds from the microscopic update rule to a group-level mean-field description, then to the linear instability thresholds of the active disordered state, and finally to the non-linear existence and stability boundaries of the balanced modular branch.
% ------------------------------------------------------------
\section{Well-mixed BChS mean-field derivation}
\label{appsec:wellmixed_bchs}

\subsection*{Microscopic update and one-step transitions}

Each node carries $o_i(t)\in\{-1,0,+1\}$. One update selects an edge $(i,j)$ uniformly at random, then chooses a direction $(a,b)\in\{(i,j),(j,i)\}$ with probability $1/2$, where $a$ is the updated node and $b$ is its neighbour. Draw
\begin{equation}
\mu=
\begin{cases}
+1, & \text{with probability } 1-p,\\
-1, & \text{with probability } p,
\end{cases}
\end{equation}
and update
\begin{equation}
o_a(t+1)=\mathrm{clip}\!\big(o_a(t)+\mu\,o_b(t)\big),\qquad \mathrm{clip}(x)\in\{-1,0,+1\}.
\label{eq:app_bchs_update}
\end{equation}

For $o_b=0$, no change occurs. For $o_b=\pm1$, the one-step transitions are:
\begin{itemize}
\item If $o_b=+1$: for $\mu=+1$, $-1\to0$, $0\to+1$, $+1\to+1$; for $\mu=-1$, $+1\to0$, $0\to-1$, $-1\to-1$.
\item If $o_b=-1$: for $\mu=+1$, $+1\to0$, $0\to-1$, $-1\to-1$; for $\mu=-1$, $-1\to0$, $0\to+1$, $+1\to+1$.
\end{itemize}
There is no direct $+1\to-1$ or $-1\to+1$ transition in one step.

\subsection*{Fractions and gain--loss equations}

Let $f_+(t),f_-(t),f_0(t)$ be the fractions of $+1,-1,0$, with $f_++f_-+f_0=1$. Define
\begin{equation}
m\equiv f_+-f_- \quad \text{(global average opinion)},
\qquad
s\equiv f_++f_- \quad \text{(non-neutral fraction)},
\qquad
f_0=1-s.
\label{eq:app_ms_onegroup}
\end{equation}
In the well-mixed closure, the neighbour state distribution equals the population fractions:
\[
P_+=f_+,\qquad P_-=f_-,\qquad P_0=f_0.
\]
From the transition table:
\begin{align}
\dot f_+
&=
f_0\Big[(1-p)P_+ + pP_-\Big]
-
f_+\Big[pP_+ + (1-p)P_-\Big],
\label{eq:app_fplus_gainloss}\\
\dot f_-
&=
f_0\Big[(1-p)P_- + pP_+\Big]
-
f_-\Big[pP_- + (1-p)P_+\Big].
\label{eq:app_fminus_gainloss}
\end{align}

\subsection*{Closed one-group ODEs in $(m,s)$ (intermediate steps shown)}

Use
\[
f_+=\frac{s+m}{2},\qquad f_-=\frac{s-m}{2},\qquad f_0=1-s,\qquad P_\pm=f_\pm.
\]

\paragraph*{Global average opinion.}
Compute $\dot m=\dot f_+-\dot f_-$:
\begin{align*}
\dot m
&= f_0\Big[(1-p)P_+ + pP_- - (1-p)P_- - pP_+\Big]
      -\Big[f_+(pP_+ + (1-p)P_-) - f_-(pP_- + (1-p)P_+)\Big]\\
&= f_0(1-2p)(P_+-P_-)
   -\Big[p(f_+P_+-f_-P_-)+(1-p)(f_+P_- - f_-P_+)\Big].
\end{align*}
With $P_\pm=f_\pm$, $P_+-P_-=m$. Next,
\[
f_+P_+-f_-P_- = f_+^2-f_-^2 = (f_+-f_-)(f_++f_-)=ms,
\qquad
f_+P_- - f_-P_+=0,
\]
so
\[
\dot m = (1-s)(1-2p)m - psm.
\]
This simplifies to
\begin{equation}
\dot m = \left[(1-2p)\Big(1-\frac{s}{2}\Big)-\frac{s}{2}\right]m.
\label{eq:app_mdot_onegroup}
\end{equation}

\paragraph*{Non-neutral fraction.}
Compute $\dot s=\dot f_+ + \dot f_-$:
\begin{align*}
\dot s
&= f_0(P_+ + P_-)
   -\Big[p(f_+P_+ + f_-P_-) + (1-p)(f_+P_- + f_-P_+)\Big]\\
&= (1-s)s-\left[p(f_+^2+f_-^2) + (1-p)2f_+f_-\right].
\end{align*}
Using
\[
f_+^2+f_-^2=\frac{s^2+m^2}{2},
\qquad
2f_+f_-=\frac{s^2-m^2}{2},
\]
gives
\begin{equation}
\dot s = s - \frac{3}{2}s^2 + \Big(\frac{1}{2}-p\Big)m^2.
\label{eq:app_sdot_onegroup}
\end{equation}

\subsection*{Fixed points and threshold}

The disordered fixed point with nonzero activity is obtained by setting $m=0$ and $\dot s=0$:
\begin{equation}
m^*=0,\qquad s^*=\frac{2}{3}
\qquad \Longleftrightarrow \qquad
f_+^*=f_-^*=f_0^*=\frac{1}{3}.
\label{eq:app_disordered_fp_onegroup}
\end{equation}
Linearising \eqref{eq:app_mdot_onegroup} at $(m,s)=(0,2/3)$ gives
\[
\dot m \approx \frac{1-4p}{3}m,
\]
hence
\begin{equation}
p_c=\frac{1}{4}.
\label{eq:app_pc_onegroup}
\end{equation}

\subsection*{Ordered branch}

For $m\neq 0$, stationarity $\dot m=0$ in \eqref{eq:app_mdot_onegroup} gives
\begin{equation}
 (1-2p)\Big(1-\frac{s}{2}\Big)-\frac{s}{2}=0
\quad\Longrightarrow\quad
s^+=\frac{1-2p}{1-p}.
\label{eq:app_splus_onegroup}
\end{equation}
Substitute $s=s^+$ into $\dot s=0$ in \eqref{eq:app_sdot_onegroup}:
\[
0=s-\frac{3}{2}s^2+\Big(\frac12-p\Big)m^2
\quad\Longrightarrow\quad
m^2=\frac{s(3s-2)}{1-2p}.
\]
Using $s^+=(1-2p)/(1-p)$ yields
\begin{equation}
(m^+)^2=\frac{1-4p}{(1-p)^2}.
\label{eq:app_mplus_onegroup}
\end{equation}
Thus the ordered branch exists for $p<1/4$.

% ============================================================
\section{Multigroup SBM mean-field to planted-partition spectrum}
\label{appsec:multigroup_sbm}

\subsection*{Group fractions and group mixing matrix}

Partition the network into $c$ groups. In group $g$, define fractions
\[
f_+^{(g)},\ f_-^{(g)},\ f_0^{(g)},\qquad f_+^{(g)}+f_-^{(g)}+f_0^{(g)}=1,
\]
and
\begin{equation}
m_g\equiv f_+^{(g)}-f_-^{(g)} \quad \text{(group-level average opinion)},
\qquad
s_g\equiv f_+^{(g)}+f_-^{(g)} \quad \text{(non-neutral fraction)},
\qquad
f_0^{(g)}=1-s_g.
\label{eq:app_ms_groups}
\end{equation}

Let $\Pi=[\pi_{gh}]$ be the group mixing matrix:
\begin{equation}
\pi_{gh}\equiv \Pr(\text{a random neighbour of a node in group $g$ lies in group $h$}),
\qquad
\sum_{h=1}^c \pi_{gh}=1.
\label{eq:app_Pi_def}
\end{equation}

Conditioned on the updated node being in group $g$, the neighbour distribution is
\begin{equation}
P_+^{(g)}=\sum_{h=1}^c \pi_{gh} f_+^{(h)},
\qquad
P_-^{(g)}=\sum_{h=1}^c \pi_{gh} f_-^{(h)},
\qquad
P_0^{(g)}=\sum_{h=1}^c \pi_{gh} f_0^{(h)}.
\label{eq:app_Pmix_groups}
\end{equation}

Define mixed fields
\begin{equation}
S_g \equiv \sum_{h=1}^c \pi_{gh} s_h,
\qquad
M_g \equiv \sum_{h=1}^c \pi_{gh} m_h,
\label{eq:app_SM_defs}
\end{equation}
so that
\begin{equation}
P_+^{(g)}=\frac{S_g+M_g}{2},
\qquad
P_-^{(g)}=\frac{S_g-M_g}{2},
\qquad
P_0^{(g)}=1-S_g.
\label{eq:app_Ppm_SM}
\end{equation}

\subsection*{Group-wise gain--loss}

For each $g$,
\begin{align}
\dot f_+^{(g)}
&=
f_0^{(g)}\Big[(1-p)P_+^{(g)} + p\,P_-^{(g)}\Big]
-
f_+^{(g)}\Big[p\,P_+^{(g)} + (1-p)P_-^{(g)}\Big],
\label{eq:app_fplus_group}\\
\dot f_-^{(g)}
&=
f_0^{(g)}\Big[(1-p)P_-^{(g)} + p\,P_+^{(g)}\Big]
-
f_-^{(g)}\Big[p\,P_-^{(g)} + (1-p)P_+^{(g)}\Big].
\label{eq:app_fminus_group}
\end{align}

\subsection*{Closed ODEs for $(m_g,s_g)$ (intermediate steps shown)}

Suppress $(g)$ temporarily. Using $\dot m=\dot f_+-\dot f_-$:
\begin{align*}
\dot m
&= f_0\Big[(1-p)P_+ + pP_- - (1-p)P_- - pP_+\Big]
   -\Big[f_+(pP_+ + (1-p)P_-) - f_-(pP_- + (1-p)P_+)\Big]\\
&= f_0(1-2p)(P_+-P_-)-(L_+-L_-),
\end{align*}
where
\[
L_+-L_-=
f_+(pP_+ + (1-p)P_-)-f_-(pP_- + (1-p)P_+).
\]
Using $P_\pm=\tfrac12(S\pm M)$ gives
\[
pP_+ + (1-p)P_-=\frac{S-(1-2p)M}{2},
\qquad
pP_- + (1-p)P_+=\frac{S+(1-2p)M}{2},
\]
hence
\begin{align*}
L_+-L_-
&=\frac12\Big[f_+(S-(1-2p)M)-f_-(S+(1-2p)M)\Big]\\
&=\frac12\Big[(f_+-f_-)S-(1-2p)(f_++f_-)M\Big]
=\frac12\Big[m_g S_g-(1-2p)s_g M_g\Big].
\end{align*}
Insert $P_+-P_-=M_g$ and $f_0=1-s_g$:
\begin{align*}
\dot m_g
&=(1-s_g)(1-2p)M_g-\frac12 m_g S_g+\frac12(1-2p)s_g M_g\\
&=(1-2p)\Big(1-\frac{s_g}{2}\Big)M_g-\frac{S_g}{2}m_g.
\end{align*}
Therefore
\begin{equation}
\dot m_g
=
(1-2p)\Big(1-\frac{s_g}{2}\Big)M_g
-\frac{S_g}{2}\,m_g.
\label{eq:app_mgdot_final}
\end{equation}

Similarly, using $\dot s=\dot f_+ + \dot f_-$:
\begin{align*}
\dot s
&= f_0(P_+ + P_-)
   -\Big[f_+(pP_+ + (1-p)P_-) + f_-(pP_- + (1-p)P_+)\Big]\\
&= f_0 S - (L_+ + L_-),
\end{align*}
and with $P_\pm=\tfrac12(S\pm M)$,
\begin{align*}
L_+ + L_-
&=\frac12\Big[f_+(S-(1-2p)M)+f_-(S+(1-2p)M)\Big]\\
&=\frac12\Big[(f_++f_-)S+(1-2p)(f_- - f_+)M\Big]
=\frac12\Big[s_g S_g-(1-2p)m_g M_g\Big].
\end{align*}
Hence
\begin{align*}
\dot s_g
&=(1-s_g)S_g-\frac12 s_g S_g+\frac{1-2p}{2}m_g M_g
= S_g\Big(1-\frac32 s_g\Big)+\frac{1-2p}{2}m_g M_g.
\end{align*}
Therefore
\begin{equation}
\dot s_g
=
S_g\Big(1-\frac{3}{2}s_g\Big)
+\frac{1-2p}{2}\,m_g\,M_g.
\label{eq:app_sgdot_final}
\end{equation}

\subsection*{Planted-partition parametrisation in terms of $a$ and $c$}

Assume an equal-size planted-partition structure: a neighbour of a node in group $g$ lies in the same group with probability $a$, and in each other group with equal probability $b$. Then
\begin{equation}
\pi_{gg}=a,\qquad \pi_{g\neq h}=b,\qquad a+(c-1)b=1,
\qquad
b=\frac{1-a}{c-1}.
\label{eq:app_ab_def}
\end{equation}
Equivalently,
\begin{equation}
\Pi=(a-b)I+b\,\mathbf 1\mathbf 1^T,
\label{eq:app_Pi_rank1}
\end{equation}
where $\mathbf 1=(1,\dots,1)^T$.

\subsection*{Spectrum of $\Pi$ (uniform and modular modes)}

From row-stochasticity, $\Pi\mathbf 1=\mathbf 1$, so the uniform mode has eigenvalue $1$.
If $\mathbf x$ satisfies $\mathbf 1^T\mathbf x=0$, then $\mathbf 1\mathbf 1^T\mathbf x=\mathbf 0$ and
\[
\Pi\mathbf x=(a-b)\mathbf x.
\]
Thus the modular eigenvalue is
\begin{equation}
\lambda \equiv \lambda_{\mathrm{mod}}=a-b=\frac{ac-1}{c-1}.
\label{eq:app_lambda_mod_ac}
\end{equation}
% ============================================================
\section{Two-group case and two-supercluster reduction (existence and stability boundaries)}
\label{appsec:twogroup_supercluster}

\subsection*{Two-group mixing and symmetric/modular cases}

For $c=2$, $b=1-a$ and
\[
\Pi=\begin{pmatrix}a&1-a\\[2pt]1-a&a\end{pmatrix},
\qquad
\lambda=2a-1.
\]
Define $M_1=a m_1+(1-a)m_2$, $M_2=(1-a)m_1+a m_2$, and similarly for $S_1,S_2$.

\paragraph*{Symmetric case (global ordering).}
If $m_1=m_2=m$ and $s_1=s_2=s$, then $M_1=M_2=m$ and $S_1=S_2=s$, giving the well-mixed system \eqref{eq:app_mdot_onegroup}--\eqref{eq:app_sdot_onegroup}. The ordered fixed point is \eqref{eq:app_splus_onegroup}--\eqref{eq:app_mplus_onegroup}.

\paragraph*{Modular case (two groups oppose).}
Assume
\begin{equation}
m_2=-m_1,\qquad s_2=s_1=s.
\label{eq:app_modular_two_group}
\end{equation}
Then $S_1=S_2=s$ and
\[
M_1=\lambda m_1,\qquad M_2=\lambda m_2.
\]
Equation \eqref{eq:app_mgdot_final} for $m_1$ becomes
\begin{equation}
\dot m_1=\left[(1-2p)\lambda\Big(1-\frac{s}{2}\Big)-\frac{s}{2}\right]m_1.
\label{eq:app_mdot_modular2}
\end{equation}

\subsection*{Balanced modular branch: existence boundary}

Introduce
\begin{equation}
q\equiv (1-2p)\lambda.
\label{eq:app_q_def}
\end{equation}
For a nontrivial modular fixed point ($m_1\neq 0$), the bracket in \eqref{eq:app_mdot_modular2} must vanish:
\begin{equation}
q\Big(1-\frac{s}{2}\Big)-\frac{s}{2}=0
\quad\Longrightarrow\quad
s=\frac{2q}{1+q}.
\label{eq:app_s_minus}
\end{equation}
On the same reduction, \eqref{eq:app_sgdot_final} gives
\begin{equation}
\dot s = s\Big(1-\frac{3}{2}s\Big)+\frac{q}{2}m_1^2.
\label{eq:app_sdot_modular2}
\end{equation}
Impose $\dot s=0$ and substitute $s=\frac{2q}{1+q}$:
\begin{align*}
0
&= \frac{2q}{1+q}\left(1-\frac{3}{2}\frac{2q}{1+q}\right)+\frac{q}{2}m_1^2\\
&= \frac{2q(1-2q)}{(1+q)^2}+\frac{q}{2}m_1^2,
\end{align*}
so for $q\neq 0$,
\begin{equation}
m_1^2=\frac{4(2q-1)}{(1+q)^2}.
\label{eq:app_m_minus_amp}
\end{equation}
Therefore the modular branch exists iff
\begin{equation}
q>\frac12
\qquad\Longleftrightarrow\qquad
\lambda(1-2p)>\frac12.
\label{eq:app_modular_exist}
\end{equation}
Equivalently,
\begin{equation}
p_{\mathrm{exist}}=\frac12\left(1-\frac{1}{2\lambda}\right),
\label{eq:app_green_line}
\end{equation}
with $\lambda=2a-1$ for $c=2$.

\subsection*{Two-supercluster reduction for general even $c$}

For general even $c$, partition groups into two sets $C_+$ and $C_-$ of equal size $c/2$ and assume
\[
m_g=
\begin{cases}
+m, & g\in C_+,\\
-m, & g\in C_-,
\end{cases}
\qquad
s_g=s\ \ \forall g.
\]
For planted-partition mixing with parameters $(a,c)$, the modular eigenvalue is
\[
\lambda=a-b=\frac{ac-1}{c-1}.
\]
On this reduction the mean-field equations collapse to the two-group modular case with $\lambda$ replaced by $\frac{ac-1}{c-1}$. Hence \eqref{eq:app_s_minus}--\eqref{eq:app_green_line} remain valid with $\lambda=\frac{ac-1}{c-1}$.

\subsection*{Stability boundary (drift toward global consensus)}

Let the two-supercluster variables be $(m_+,s_+)$ and $(m_-,s_-)$ with base state
\[
(m_+,s_+,m_-,s_-)=(m,s,-m,s),
\]
where $(m,s)$ satisfy \eqref{eq:app_s_minus}--\eqref{eq:app_m_minus_amp} with $q=\lambda(1-2p)$.

Use perturbations
\[
\delta m_+=\delta m_-\equiv x,
\qquad
\delta s_+=-\delta s_-\equiv y.
\]
These satisfy $\delta M_+=\delta M_-=x$ (uniform mode, eigenvalue $1$) and $\delta S_+=\lambda y$, $\delta S_-=-\lambda y$ (modular mode, eigenvalue $\lambda$).

Linearising the two-supercluster mean-field equations about the base state gives
\begin{equation}
\begin{pmatrix}\dot x\\[2pt]\dot y\end{pmatrix}
=
\begin{pmatrix}
a_{11} & a_{12}\\
a_{21} & a_{22}
\end{pmatrix}
\begin{pmatrix}x\\[2pt]y\end{pmatrix},
\label{eq:app_xy_system}
\end{equation}
with
\begin{align}
a_{11} &= (1-2p)\left(1-\frac{s}{2}\right)-\frac{s}{2},
\label{eq:app_a11}\\
a_{12} &= -\,m\,\lambda(1-p),
\label{eq:app_a12}\\
a_{21} &= \frac{1-2p}{2}\,m\,(\lambda+1),
\label{eq:app_a21}\\
a_{22} &= \lambda\left(1-\frac{3}{2}s\right)-\frac{3}{2}s.
\label{eq:app_a22}
\end{align}

Stability loss occurs when an eigenvalue crosses zero:
\begin{equation}
a_{11}a_{22}-a_{12}a_{21}=0.
\label{eq:app_det_zero}
\end{equation}
Substitute $s=\frac{2q}{1+q}$ and $m^2=\frac{4(2q-1)}{(1+q)^2}$ with $q=\lambda(1-2p)$. The determinant condition reduces to
\begin{equation}
(1+\lambda)q^2 + (2\lambda^2+\lambda-2)q - \lambda^2 = 0.
\label{eq:app_q_quadratic}
\end{equation}
Let $q_c(\lambda)$ be the positive root of \eqref{eq:app_q_quadratic}. Since $q=\lambda(1-2p)$, the stability boundary is
\begin{equation}
p_{\mathrm{stable}}=\frac12\left(1-\frac{q_c(\lambda)}{\lambda}\right),
\label{eq:app_blue_line}
\end{equation}
with $\lambda=\frac{ac-1}{c-1}$.

\subsection*{Summary (in terms of $a$ and $c$)}

For planted-partition mixing with parameters $(a,c)$ and $b=(1-a)/(c-1)$:
\[
\lambda=\frac{ac-1}{c-1}.
\]
Existence boundary:
\[
p_{\mathrm{exist}}=\frac12\left(1-\frac{1}{2\lambda}\right).
\]
Stability boundary:
\[
p_{\mathrm{stable}}=\frac12\left(1-\frac{q_c(\lambda)}{\lambda}\right),
\]
where $q_c(\lambda)$ is the positive root of \eqref{eq:app_q_quadratic}.

The region between the two boundaries is the regime where the balanced modular branch exists but is not yet robust against consensus-forming perturbations, which is consistent with multi-attractor behaviour observed in numerical integration and Monte Carlo simulations.

\section{Robustness checks from Monte Carlo simulations}
\label{appsec:mc_robustness_bchs}

This section collects supplementary Monte Carlo results used as basic robustness checks for the phase behaviour discussed in main text. The figures are included only to show that the separation between global order $O$ and intra-group order $O_{\mathrm{intra}}$ persists under variations in network connectivity and the number of groups.

\begin{figure*}[htbp]
    \centering
    \includegraphics[width=0.98\linewidth]{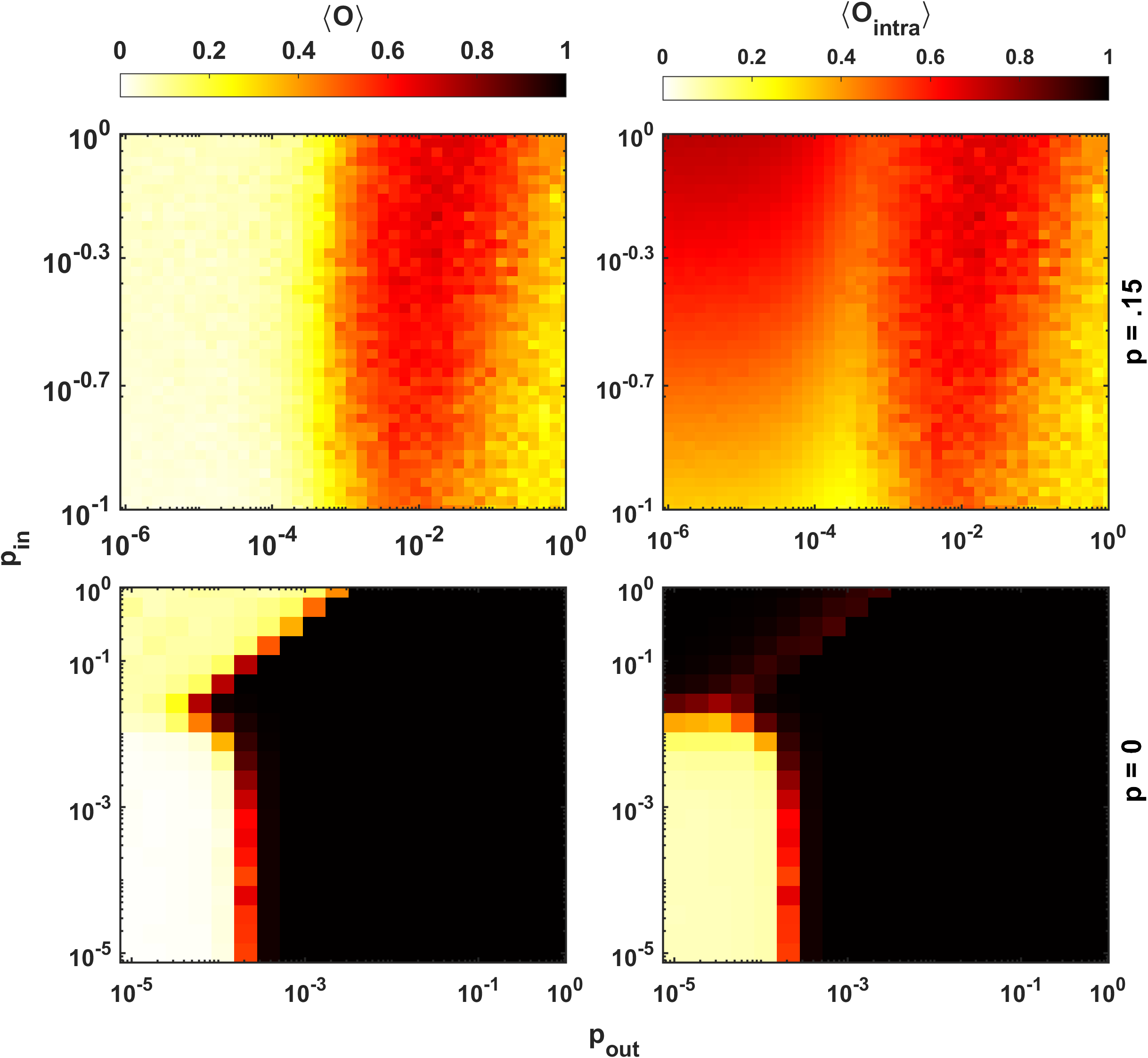}
    \caption[Phase diagram in $(p_{\mathrm{out}},p_{\mathrm{in}})$]{Global order parameter $O$ (left column) and intra-group order parameter $O_{\mathrm{intra}}$ (right column) in the $(p_{\mathrm{out}},p_{\mathrm{in}})$ plane for modular networks with $n = 10^{4}$ nodes partitioned into $c = 100$ equal groups. In the top row the disagreement probability is fixed at $p = 0.15$; both $p_{\mathrm{out}}$ and $p_{\mathrm{in}}$ are sampled on a $42 \times 42$ logarithmic grid spanning $10^{-6} \le p_{\mathrm{out}} \le 10^{0}$ and $10^{-6} \le p_{\mathrm{in}} \le 10^{0}$, and, for each parameter pair, observables are averaged over $nI = 30$ network realisations after a transient of $T_{\mathrm{s}} = 2\times 10^{6}$ steps and a measurement window of $T_{\mathrm{m}} = 2\times 10^{6}$ steps. In the bottom row the same procedure is repeated for $p = 0$ on a $20 \times 20$ logarithmic grid with $10^{-5} \le p_{\mathrm{out}} \le 10^{0}$ and $10^{-5} \le p_{\mathrm{in}} \le 10^{0}$, again using $nI = 30$ realisations per parameter pair. Colour encodes the time-averaged value of the corresponding order parameter.}
    \label{fig3:phase_pout_pin}
\end{figure*}

Figure~\ref{fig3:phase_pout_pin} shows $O$ and $O_{\mathrm{intra}}$ in the $(p_{\mathrm{out}},p_{\mathrm{in}})$ plane for $c=100$ at two representative values of $p$ ($p=0.15$ and $p=0$). The purpose of this scan is to check that the modularly ordered regime (large $O_{\mathrm{intra}}$ and small $O$) appears when disagreement is present ($p>0$) and the network is sufficiently modular (small $p_{\mathrm{out}}$, large $p_{\mathrm{in}}$).

\begin{figure*}[htbp]
    \centering
    \includegraphics[width=0.98\linewidth]{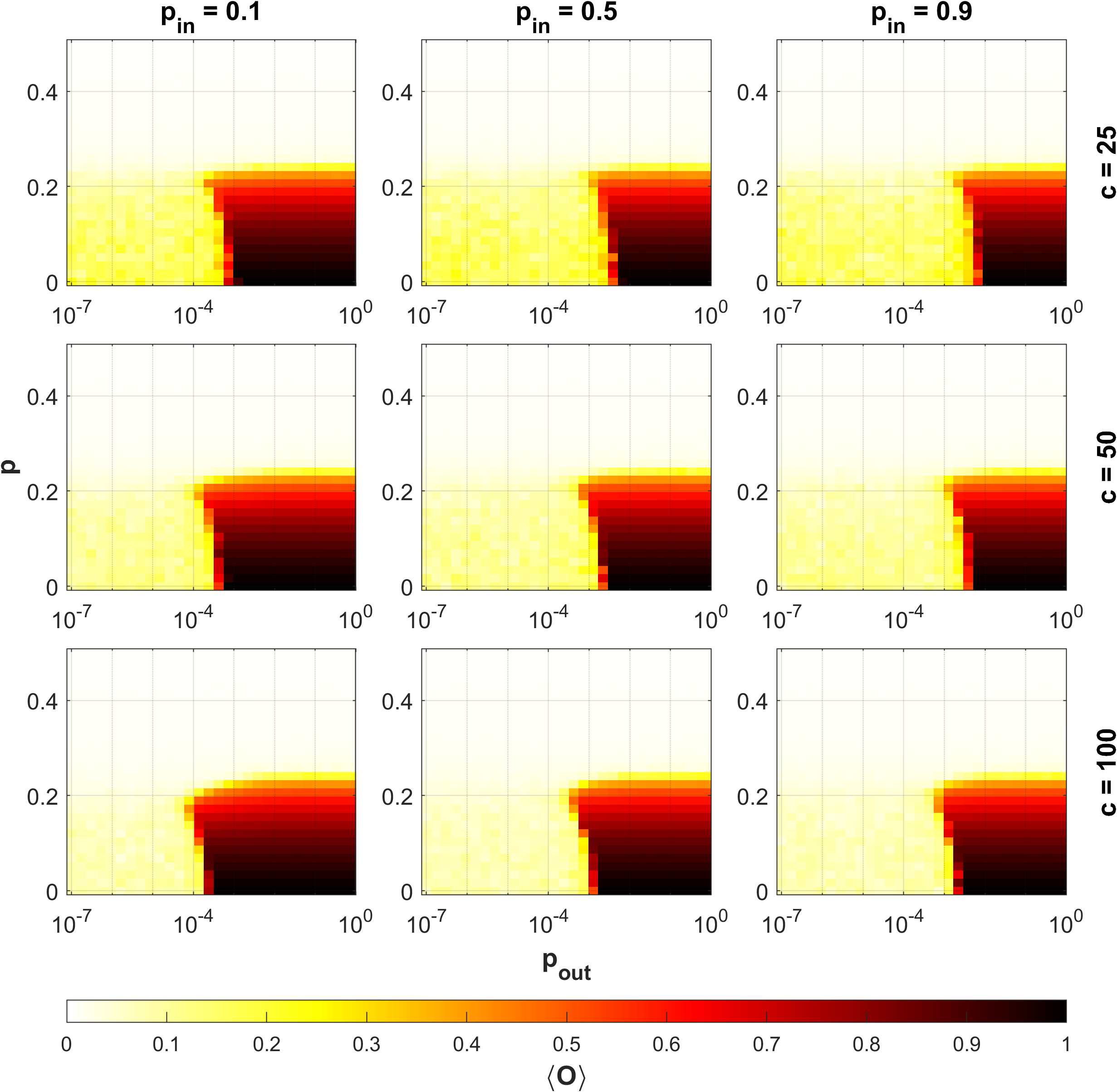}
    \caption[Global order across $c$ and $p_{\mathrm{in}}$]{Global order parameter $\langle O \rangle$ in the $(p_{\mathrm{out}},p)$ plane for modular networks, shown for different values of the number of groups $c$ (rows) and within-group link probability $p_{\mathrm{in}}$ (columns). Each panel corresponds to a stochastic block network with $n = 10^{4}$ nodes. The disagreement probability $p$ is sampled from $0$ to $0.5$, and $p_{\mathrm{out}}$ is logarithmically spaced from $10^{-7}$ to $10^{0}$ on a $30\times 30$ grid. For each parameter set, observables are averaged over a transient window ($T_{\mathrm{s}} = 2\times10^{6}$), a measurement window ($T_{\mathrm{m}} = 2\times10^{6}$), and $nI = 20$ independent network realisations. Colour encodes the mean global order parameter $\langle O \rangle$.}
    \label{fig4:gridC}
\end{figure*}

\begin{figure*}[htbp]
    \centering
    \includegraphics[width=0.98\linewidth]{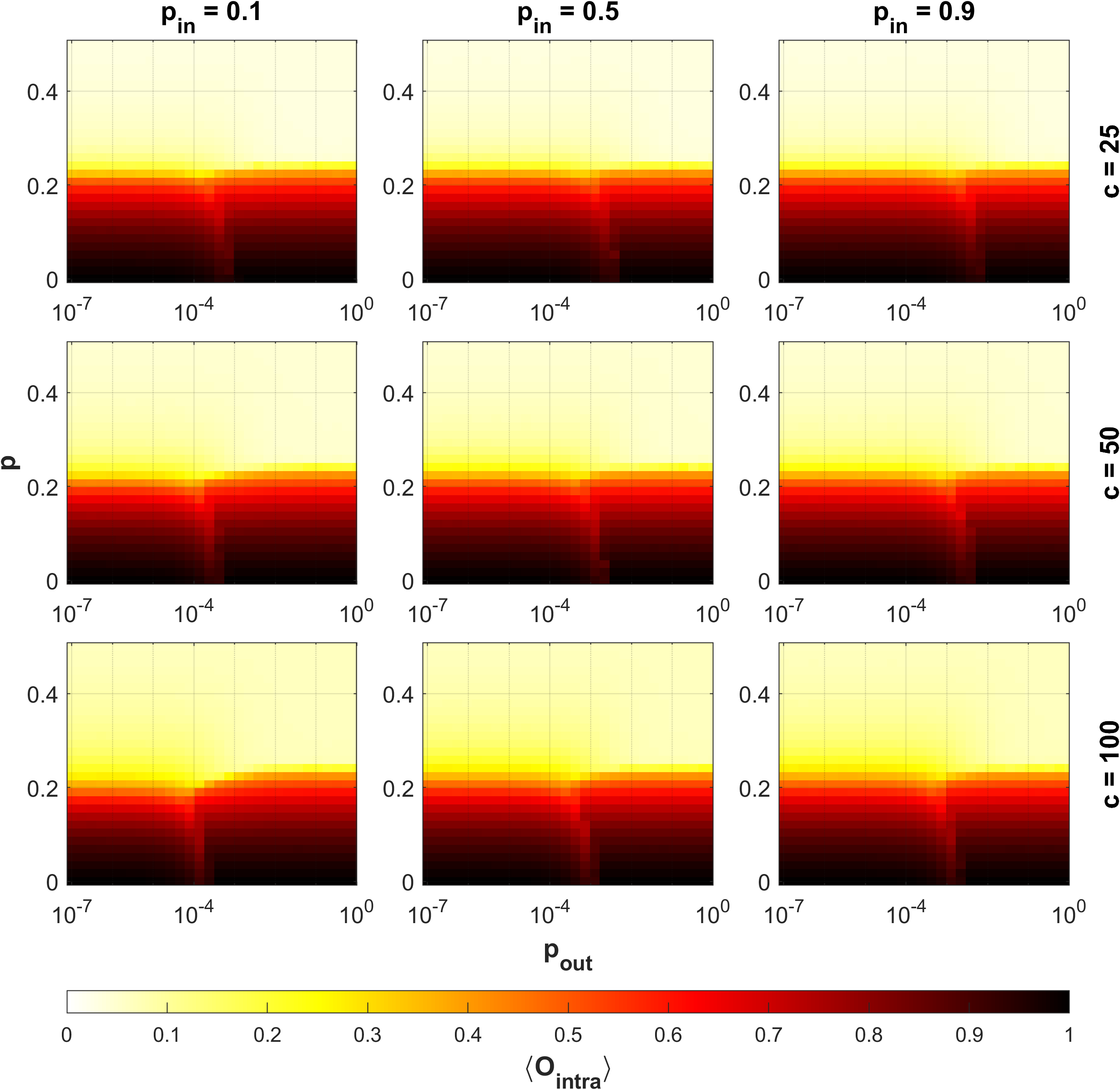}
    \caption[Intra-group order across $c$ and $p_{\mathrm{in}}$]{Intra-group order parameter $\langle O_{\mathrm{intra}} \rangle$ in the $(p_{\mathrm{out}},p)$ plane for modular networks, shown for different values of the number of groups $c$ (rows) and within-group link probability $p_{\mathrm{in}}$ (columns). The simulation grid, averaging windows, and ensemble size are the same as in Figure~\ref{fig4:gridC}. Colour encodes the mean intra-group order parameter $\langle O_{\mathrm{intra}} \rangle$.}
    \label{fig5:gridCintra}
\end{figure*}

Figures~\ref{fig4:gridC} and \ref{fig5:gridCintra} show supplementary $(p_{\mathrm{out}},p)$ scans for different values of $c$ and $p_{\mathrm{in}}$. These are included as basic robustness checks to show that:
\begin{itemize}
\item the loss of global order $O$ occurs at lower disagreement levels than the loss of intra-group order $O_{\mathrm{intra}}$, and
\item this separation persists across different numbers of groups and different within-group connectivities.
\end{itemize}
This confirms that the modularly ordered regime is not restricted to a single choice of $(c,p_{\mathrm{in}})$.

\bibliography{MyRefs}
\end{document}